\newcommand{\beq}{  \begin{eqnarray}}
\newcommand{\eeq}{  \end{eqnarray}}

\documentclass[pra,twocolumn,showpacs,preprintnumbers,amsmath,amssymb]{revtex4-1}

\usepackage{graphicx}% Include figure files
\usepackage{dcolumn}% Align table columns on decimal point
\usepackage{bm}% bold math

\begin{document}
    	
\title{Geometric gauge potentials and forces in low-dimensional scattering systems}
    
\author{B. Zygelman}
\email{bernard@physics.unlv.edu}
\affiliation{%
Department of Physics and Astronomy, University of Nevada, Las Vegas, Las Vegas NV 89154
}%

\date{\today}
\begin{abstract}
We introduce and analyze several low-dimensional scattering systems that exhibit geometric
phase phenomena. The systems are fully solvable and we compare accurate solutions of them with
those obtained in a Born-Oppenheimer projection approximation. We 
illustrate how geometric magnetism manifests in them, and explore the relationship between
solutions obtained in the diabatic and adiabatic pictures. 
We provide an example, involving a neutral atom dressed by an external field, in which the system
mimics the behavior of a charged particle that interacts with, and is scattered by, a ferromagnetic material.
We also introduce a similar system that exhibits Aharonov-Bohm scattering. We propose possible practical applications.
We provide a theoretical approach that underscores universality 
in the appearance of geometric gauge forces. We do not insist on degeneracies
in the adiabatic Hamiltonian, and we posit that the emergence of geometric gauge forces is
a consequence of symmetry breaking in the latter. 
\end{abstract}

\maketitle
\section{Introduction}
The Born-Oppenheimer (BO) approximation allows the replacement of a
complex many-body problem by a mean field description
in which the motion of atoms can be accurately described. The dynamics of the latter 
are governed by an effective force ${\bm F}$ generated by a scalar potential that represents the collective 
motion of the fast degrees of freedom, i.e. the electrons. According to the Hellmann-Feynman theorem\cite{feyn39} 
\beq
{\bm F} = - {\bm \nabla} \, \epsilon({\bm R})  
\label{0.1}
\eeq
where $ \epsilon $ is the Born-Oppenheimer energy (including higher-order corrections\cite{dal56}) of the electronic state and ${\bm R}$ typically
describes the geometry of the atomic ensemble. Molecular spectroscopy, atomic
collision phenomena, as well as molecular dynamics simulations are just some of the applications that derive
their utility from the implications of Eq. (\ref{0.1}).
In its modern expression\cite{moo86,shap89,zyg87a} BO theory 
also allows for an additional contribution to Eq. (\ref{0.1}), that of an
induced effective ``magnetic'' force\cite{zyg87a,ber89,zyg90,bli05} 
\beq
&& {\bm F}_{m} \equiv  \frac{{\bm v}}{2} \times {\bm B} - {\bm B} \times \frac{{\bm v}}{2}   \nonumber \\
&& {\bm B} = {\bm \nabla} \times \hbar {\bm A}   
\label{0.2}
\eeq
where ${\bm v}$ is the atom velocity operator\cite{zyg87a} and $ {\bm A}$ is a vector potential. This effective force
is sometimes called geometric magnetism, a term coined by M.V. Berry,
and is the nomenclature we adopt here. 

Advances\cite{lin09a,lin09b,simon11,spiel11,dal10b,dal10a} in the manipulation of cold atomic matter
has allowed researchers to ``dress'' atoms via the application of laser fields in order to
induce gauge potentials that manifest as effective Lorentz forces on the atoms.
This phenomenon has been called, ``artificial''\cite{dal10b} or ``synthetic''\cite{lin09b} magnetism and 
in a laboratory realization of it\cite{lin09a}, a BEC cloud was observed to undergo cyclotron motion in the same manner  
charged particles behave in a magnetic field. The effect has important implications for the control of atomic matter, 
and offers a novel route to quantum magnetism\cite{spiel11,diaz11}.

In this paper we address the question; what is the relationship between geometric and induced synthetic magnetism?
The appearance of vector gauge potentials in molecular physics was first demonstrated by Mead and Truhlar\cite{mead76} 
in their study of pathologies associated with conical intersections (degeneracies of distinct BO molecular energies at an isolated
point ${\bm R}_{c}$). At a conical intersection the phase of the BO wavefunction is undefined and, in order to avoid multi-valuedness,
Mead and Truhlar introduced a compensating phase factor to accompany the nuclear wavefunction. They argued that such a  procedure is equivalent to
minimally coupling the nuclear motion with
a non-trivial vector
potential\cite{mead76}. The Mead-Truhlar gauge potential does not
exhibit geometric magnetism but since Berry's vector potential\cite{ber84,ber89} has non-vanishing curl, 
systems exhibiting the latter do. In the Mead-Truhlar treatment degeneracy plays a central role as it does in Berry's Hamiltonian\cite{ber84}. 
Degeneracy is also an essential ingredient in the non-Abelian generalization of Berry's phase\cite{zee}. 
Thus the appearance of geometric
phases has historically been associated with systems in which degeneracies manifest. This does not appear to be the case for artificial
magnetism\cite{dal10b} and so it is worthwhile to further investigate the relationship between the two phenomena.
     
In applications\cite{zyg92,zyg94a,zyg09} it is often necessary to go beyond BO theory and exploit
a multi-channel generalization of it, the Born-Huang expansion or the method of perturbed stationary states (PSS)\cite{mott49}.
In atomic collision theory the latter lead to equations that have the form\cite{zyg87a} 
\beq
-\frac{\hbar^{2}}{2 m} ({\bm \nabla} - i {\bm  A})^{2} {F} + { V} \, { F} =E { F} 
\label{0.3}
\eeq
where ${ F}$ is a multi-channel amplitude expressed as a column matrix, ${\bm { A}}$ is a 
matrix-valued vector potential, and $ { V}$ is a diagonal scalar-valued matrix whose elements consist of the
BO eigenvalues associated with each channel and $E$ is the total energy eigenvalue. If a complete set of channel states are included in the PSS expansion then Eq. (\ref{0.3})
is exact. We distinguish two cases in applications of Eq. (\ref{0.3}). In the first, matrix $ { V}$ is degenerate at all
${\bm R}$ and a gauge transformation can be found\cite{zyg87a,zyg90}  so that, in the new gauge, ${\bm A}=0$. 
The representation in which ${\bm A}=0$ is sometimes called the diabatic picture\cite{Smith69,zyg87a}, whereas Eq. (\ref{0.3}) describes
dynamics in the adiabatic picture. 
The
second case, which we deal with exclusively in this paper, ${ V}$ is non-degenerate and Eq. (\ref{0.3})
exhibits gauge covariance if the vector valued potential ${\bm A}$ is taken as the spatial component of a 3+1 gauge field
that also contains a temporal component\cite{zyg87a}. 
In applications\cite{zyg92,zyg94a,zyg09} one typically 
truncates the PSS equations to a finite number of channels (most realistic systems involve an infinite number of channels)
and in that case one cannot, in general, find a transformation into a gauge where $ {\bm A}=0$. As a consequence
 effective gauge forces\cite{zyg87a,zyg92} manifest. Does this imply that the existence of the latter are
the artifacts of an approximation procedure? What is the relationship between geometric forces  
and the fictitious forces that arise in a classical description where non-inertial frames are employed?  

In order to address these, and other questions, we
introduce and analyze several low dimensional but solvable models that exhibit geometric phase phenomena in the BO approximation. In the
systems discussed here the ``fast'' degrees of freedom span a finite Hilbert space of dimensions $d=2,4$. Similar models
have been discussed previously\cite{ber93,ber96,berry10} but here we focus on systems that do not exhibit energy degeneracies
in the adiabatic approximation. We concentrate our efforts on fully quantal treatments of Eq. (\ref{0.3}) and we
limit our discussion to scattering systems and their solution.

In section II we provide a general theoretical framework for the subsequent discussions.
We introduce a 1D scattering system in which geometric phase phenomena arise and which
fully quantal analytic solutions are available. In section III we introduce a 2D scattering system that
exhibits the phenomenon of synthetic, or geometric, magnetism. We compare the scattering solutions of it
with that obtained for a charged particle scattered by a ferromagnetic slab. In the remaining sections
we show how these systems are related to ones in which geometric forces arise due to inter-particle interactions.
An example involving two, interacting, spin-1/2 particles is presented. 

Vector valued quantities
will be shown in boldface, we will announce matrix-valued quantities in the text and use
ordinary Roman typeface to represent them. However, in certain circumstance, where there might be
a possibility for confusion, we will underline a Roman letter to stress its matrix nature.       
     
\section{} 
\subsection{General Theory}
Consider the Hamiltonian 
\beq
H = -\frac{\hbar^{2}}{2 m} \, {\bm {\nabla}_{{\bm R}}}^{2} + H_{ad}({\bm R}) 
\label{1.1}
\eeq
where $ H_{ad}({\bm R})$ is the Hamiltonian defined on an
$n$-dimensional Hilbert space $h_{n}$ that represents the internal, or ``fast'' \cite{zyg87a}, degrees of freedom of a quantum system of mass $m$.
It is parameterized by the eigenvalues of the quantum variable ${\bm R}$. In molecular physics $H_{ad}$ is called
the adiabatic Hamiltonian and represents the total kinetic, electrostatic, and magnetic interactions
among the electrons. The kinetic energy term in Eq. (\ref{1.1}) represents the motion of the ``slow'' degrees of freedom.
 In molecular physics the internal space $h_{n}$ is  spanned by  the Born-Oppenheimer (BO)
eigenstates of $H_{ad}$. Additional realizations described by Eq. (\ref{1.1}) 
could be an atom, molecule or  spin-$n$ system in an external field that is modulated
by the values of ${\bm R}$. If we assume that the basis for $h_{n}$ is finite, we can express
\beq
H_{ad}({\bm R}) =  U({\bm R}) H_{BO} U^{\dag}({\bm R}) 
\label{1.2}
\eeq
where $H_{BO}$ is a diagonal matrix whose entries we shall label $e_{1}({\bm R}), e_{2}({\bm R}), ... e_{n}({\bm R})$. 
In molecular physics they are called the BO eigenvalues for the electronic Hamiltonian $H_{ad}$.   
$U({\bm R})$ is a unitary operator acting on the internal states and, in general, is also parameterized
by ${\bm R}$. Expressed as a matrix representation in the BO basis, $ { U}$  is an $ n \times n$ unitary matrix.

Any well behaved Hermitian operator $H_{ad}$ can be written in the form given by Eq. (\ref{1.2}),
however we require the additional condition that ${ U}$ must be single-valued in the parameter space spanned by ${\bm R}$.
Consider the unitary matrix,
\beq
W^{\dag}({\bm R}) = {\cal P} \exp(i \int_{{\cal C}}^{\bm R} \, d {\bm R'} \cdot {\bm A}({\bm R}')) \, C
\label{1.3}
\eeq
where ${\bm A}$ is a vector valued $n \times n$ matrix (i.e. a non-Abelian gauge potential), 
${\cal P}$ represents a  path-ordering operator and $C$ is a constant unitary matrix. Path-ordered integrals along curve ${\cal C}$ in parameter space ${\bm R}$ are defined
as follows: Consider a path, ${\cal C}$, defined by the set $ {\bm R}_{1}, {\bm R}_{2}, ...{\bm R}_{n} $  where $
{\bm R}_{1}={\bm R}(t_{1}),{\bm R}_{2}={\bm R}(t_{2})...{\bm R}_{n} ={\bm R}(t_{n})$, ${\bm R}_{n}={\bm R}$, and $t_{1} < t_{2} <... t_{n}$
defines a trajectory that maps out the path.
We define
\beq
&& {\cal P} \exp(i \int_{{\cal C}}^{\bm R} \, d {\bm R'} \cdot {\bm A}({\bm R}') )
\equiv  \nonumber \\
&& T \exp(i \int_{t_{1}}^{t_{n}} dt \, \frac{d {\bm R}(t)}{d t} \cdot  {\bm A}(t) ) 
\label{1.4}
\eeq
where we assumed that the path is sufficiently smooth so that $\frac{d {\bm R}(t)}{d t}$ is well defined on it and
$T$ is the Dyson time-ordering operator. For open ended paths, such integrals
are also called the Wilson-line\cite{mansour2000}, and if ${\cal C}$ traces a closed path Eq. (\ref{1.3}) represents the
Wilson loop integral\cite{wuyang76} which we require, for a pure gauge, to have the value of unity. 
To that end, we demand that
\beq
F_{\mu \nu}=\partial_{u} { A}_{\nu} -\partial_{\nu} { A}_{\mu}-i [ { A}_{\mu},{ A}_{\nu}]=0
\label{1.5}
\eeq
where we have used the notation defined in \cite{zyg87a} and, here, $\mu,\nu$ are spatial indices only.
We make the assumption, if ${\bm A}$ satisfies condition Eq. (\ref{1.5}) and is not singular, then the value for 
path integral Eq. (\ref{1.3}) is independent
of ${\cal C}$. With this working assumption, we take the gradient of Eq. (\ref{1.3}) 
\beq
&& {\bm \nabla} W^{\dag}  = i \, {\bm A} \, W^{\dag} \nonumber \\
&& ( {\bm \nabla} W^{\dag}) W = i {\bm A} 
\label{1.6}
\eeq
or
\beq
&& ( {\bm \nabla} W^{\dag}) W =  -W^{\dag} {\bm \nabla} W = i {\bm A} 
\label{1.7}
\eeq
and
\beq
{\bm A}=i \, W^{\dag} {\bm \nabla} W.
\label{1.8}
\eeq
Consider a differentiable unitary operator $ U({\bm R})$ so that $ i \, U^{\dag} {\bm \nabla} U = {\bm A}$, and
therefore satisfies Eq. (\ref{1.5}) \cite{zyg90,mead92}. Since
$ U$ and $W$ are both unitary and differentiable, $ W = U \, Z$ where $ Z = U^{-1} \, W$. Inserting
the expression for $ W$ into the r.h.s. of Eq. (\ref{1.8}) we obtain
\beq
&& i \, W^{\dag} {\bm \nabla} W = {\bm A} =  Z^{\dag} \,{\bm A} \, Z + i Z^{\dag} {\bm \nabla} Z \quad \text{or} \nonumber \\
&& [Z,{\bm A}]=i {\bm \nabla} Z. \nonumber
\eeq
The unitary matrix $ Z$ that relates $U$ with $W$ must obey the above constraint equation and represents
a non-Abelian gauge transformation in which $ {\bm A} $ remains invariant (e.g. for the Abelian case it requires that Z is constant). 
If $W$ and $U$ differ, they do so 
up to an inconsequential gauge transformation. 
For example, if we replace $U$ in the adiabatic
Hamiltonian, given in Eq. (\ref{1.2}), with $W$ then Hamiltonian Eq. (\ref{1.1}) is replaced with
\beq
H' = -\frac{\hbar^{2}}{2 m} \, {\bm {\nabla}_{{\bm R}}}^{2} + W H_{BO} W^{\dag}. 
\nonumber
\eeq 
It will follow from the discussion below that both $ H$, and $H'$ lead to identical Schrodinger equations in the
adiabatic gauge since both formulations share the same vector and scalar gauge potentials.
Therefore, we can always replace the adiabatic Hamiltonian with one in which $U$ is parameterized by a Wilson line
as given by Eq. (\ref{1.3}). 
\subsection{Illustrative Example}
As an illustration,  consider the direct product of an
$h_{n}=2$ dimensional Hilbert space of a two-level, or qubit, system and a 1D 
Hilbert space for quantum variable ${\bm R}$, e.g. ${\bm R}$ is represented by the one-dimensional
variable $ -\infty <x < \infty$. We
define the gauge field  
\beq
{\bm A} =\left(
\begin{array}{cc}
 A_0 & A_1 \\
 A_1 & -A_0
\end{array}
\right)
\label{1.9} 
\eeq
where $A_{0},A_{1}$ are real constants. 
It is evident that condition Eq. (\ref{1.5}) is satisfied by this
gauge potential and we can apply Eq. (\ref{1.3}) to construct  
\beq
&& U=\left(
\begin{array}{cc}
 \cos \left( A x\right)-\frac{i {A_{0}} \sin A x}{A} & -\frac{i
   {A_{1}} \sin A x}{A} \\
 -\frac{i {A_{1}} \sin A x}{A} & \cos \left(A x\right)+\frac{i
   {A_{0}} \sin \left(A x\right)}{A}
\end{array}
\right)  \nonumber \\
&& A \equiv \sqrt{{A_{1}}^2+{A_{0}}^2}.
\label{1.10}
\eeq
We note that $U$ is single-valued and $U(x=0)$ is the identity operator. 

Suppose the internal Hamiltonian is given by
\beq
H_{ad}(x) = U_{0} H_{BO}  U_{0}^{\dag} 
\label{1.11}
\eeq
where $H_{BO}$ is an arbitrary diagonal matrix. Previous studies e.g. \cite{ber93,mead76} have focused
on systems in which $H_{BO}$ is degenerate for some value of the parameter ${\bm R_{c}}$, i.e.
$e_{1}({\bm R}_{c}) = e_{2}({\bm R}_{c})$. Here {\bf we do not allow any crossings}, indeed we take
the BO eigenergies to be constant throughout the domain of $x$. We define
\beq
&& U_{0} = \left(  \begin{array}{cc} 1 & 0 \\ 0 & 1 \end{array} \right)    \quad x<0  \nonumber \\
&& U_{0} = U  \quad x >0 
\label{1.12}
\eeq
where $ U$ is given by Eq. (\ref{1.10}). $U_{0}$ is continous in $x$ but not differentiable at $x=0$.
Choosing 
\beq
H_{BO}= \left( \begin{array}{cc}  \Delta & 0 \\ 0 & -\Delta  \end{array} \right), 
\label{1.13}
\eeq
where $ \Delta >0 $ is a constant, 
we obtain the set of coupled Schrodinger equations for the spinor eigenstates of Hamiltonian (\ref{1.1})
\beq
{\underline F}'' - \frac{2 m}{\hbar^{2}} \,  { V} \, \, {\underline F} + \frac{2 m}{\hbar^{2}}  E \, {\underline F}=0  
\label{1.14}
\eeq
where
\beq
{\underline F}\equiv \left ( \begin{array}{c} F_{c}(x) \\ F_{o}(x) \end{array} \right )
\label{1.14a}
\eeq
and 
\beq
{ V}  &=& H_{BO}=  \left( \begin{array}{cc}  \Delta & 0 \\ 0 & -\Delta  \end{array} \right)  \quad  x<0 \nonumber \\
{ V} &=& U \, H_{BO} \, U^{\dag}   \quad x \geq 0.  \label{1.15} 
\eeq
We seek scattering solutions to Eq. (\ref{1.14}) and if we set $E =-\Delta + \frac{\hbar^{2} k^{2}}{2 m} $ 
where $ \frac{\hbar^{2} k^{2}}{2 m} < 2\, \Delta $, the
excited internal BO channel is closed. 

Because $H_{BO}$ is  diagonal and constant, instead of solving Eqs.(\ref{1.14}) directly it is convenient
to transform to the adiabatic picture whose wave function is,
\beq
{\underline F}_{ad} = U_{0}^{\dag} \, {\underline F}
\label{1.16}
\eeq
and satisfies\cite{zyg87a}
\beq
({\bm \nabla} - i { A})^{2} {\underline F}_{ad} - \frac{2 m}{\hbar^{2}} \, H_{BO} {\underline F}_{ad} + \frac{2 m}{\hbar^{2}} \, E {\underline F}_{ad} =0
\label{1.17}
\eeq
where $ { A} $ is the gauge potential
\beq
{ A} &=&\left(
\begin{array}{cc}
 A_0 & A_1 \\
 A_1 & -A_0
\end{array}
\right) \quad x > 0 \nonumber \\
{ A} &=& 0  \quad x< 0. 
\label{1.18}
\eeq
  In the adiabatic picture we have replaced the off-diagonal potential matrix  ${ V}$ with the diagonal BO matrix,
for the price of gauge potentials. At collision energies where some of the excited BO channels are closed,
a common approximation in molecular and collision physics is to project the system of coupled equations (which are exact) onto the open sector. This is called
the Born-Oppenheimer (BO) or, if several channels are open, the perturbed stationary states (PSS) approximation\cite{zyg87a,mott49}. In our example 
this approximation leads to the following single channel equation\cite{zyg87a},
\beq
&& (\partial + i A_{0})^{2} F_{o}  +  k^{2}  F_{o} -  b \,  F_{o} =0  \quad x \geq 0 \nonumber \\
&& \partial^{2} F_{o}  +  k^{2}  F_{o}  =0  \quad x < 0 \nonumber \\
&& b \equiv  \sum_{k \neq i} {\bm A}_{ik} \cdot {\bm A}_{ki} = A_{1}^{2}
\label{1.19}
\eeq
The induced potential $b$ is closely related to the so-called ''B'' term, or adiabatic, correction\cite{dal56,kolos,mari98}.
A gauge transformation $ F_{o} \rightarrow \exp(-i A_{0} \, x) F_{o} $ allows us to ``eliminate'' the derivative coupling in
Eq. (\ref{1.19}) and we can proceed to solve the scattering problem. Imposing the boundary condition
that the open-channel wavefunction vanishes at $ x =L >>0$ (which is equivalent to the placement of an impenetrable barrier at $L$),
we obtain, in the asymptotic region for $ x< 0$,
\beq
&&  F_{0}(x) = \exp(i k (x-L)) + R \exp(-i k (x-L)) \nonumber \\ 
&&  R = \exp(-2 i k L) \times \nonumber \\
&& \left( -1+\frac{2 k}{k+i \sqrt{k^2-A_{1}^{2}} \cot \left(L
   \sqrt{k^2-A_{1}^{2} }\right)}   \right) \quad  k> |A_{1}| \nonumber \\
&& \text{ and for } \quad  k < |A_{1}| \nonumber \\
&& R = \exp(-2 i k L) \times \nonumber \\
&& \left( -1+\frac{2 k}{k+i \sqrt{A_{1}^{2}-k^{2}} \coth \left(L
   \sqrt{A_{1}^{2}-k^{2} }\right)}   \right).
\label{1.20}
\eeq
We can also solve the fully coupled Eqs. (\ref{1.17}) analytically, as outlined in the appendix, without resorting to
the Born-Oppenheimer approximation. That solution, in the adiabatic limit, leads to
\beq
R = -1 + 2 i k \, \Bigl (L -\frac{\tanh(A_{1} L)}{A_{1}} \Bigr ) \, + \, {\cal O}(k^{2}) + ...
\label{1.25x}
\eeq
where we have kept only the lowest order terms, as $k \rightarrow 0$, in an effective range expansion for $R$. 
This result is in harmony with that obtained in an effective range expansion of expression (\ref{1.20}) obtained using the BO approximation.

In summary,
\begin{itemize}
\item At collision energies where the excited internal state is closed and the energy
defect between the open and closed channels is large, the lowest order term in an effective range expansion 
(i.e. in the cold collision energy regime) of the Born-Oppenheimer expression for the reflection coefficient is
identical to that obtained by solution of the fully coupled equations. It depends on the value of the off-diagonal component of
the induced gauge potential (\ref{1.9}) despite the fact that the pure gauge condition 
Eq. (\ref{1.5}) is satisfied.  
\item In the BO approximation the off-diagonal gauge coupling leads to a higher order induced scalar potential,
given by the expression  for $b$ in Eq. (\ref{1.19}). In this approximation the diagonal component $A_{0}$ can simply be
``gauged'' away and physical quantities, such as the scattering length, are independent of it.
\item At collision regimes where the BO approximation is no longer appropriate \cite{goss2010}, (e.g. when $ \Delta \sim E$), the diagonal
components $A_{0}$ do affect scattering properties. This may seem counter-intuitive since we stated that a gauge transformation
can be employed to  transform these potentials away. One way of proceeding beyond BO theory is to describe
the coupling between the open and closed channel through the introduction of a non-local optical potential\cite{fesh58}. In such a treatment
Eq. (\ref{1.19}) is replaced by\cite{zyg11a}
\beq
&& (\partial + i A_{0})^{2} F_{o}  +  k^{2}  F_{o} + \frac{2 m}{\hbar^{2}} \, V_{opt}  F_{o} =0  \quad x \geq 0 \nonumber \\
&& V_{opt}F_{o} \equiv \int dx'\, V_{opt}(x,x')F_{o}(x')  
\label{1.25aa}
\eeq
where $ V_{opt}$ is a nonlocal potential. Now the transformation $ F_{o} \rightarrow \exp(-i A_{0} \, x) F_{o} $  does
eliminate the derivative coupling in Eq. (\ref{1.25aa}) but, since $V_{opt}$ is non-local, it results in
\beq
&& \int dx'\, V_{opt}(x,x')F_{o}(x')  \rightarrow \nonumber \\
&& \int dx'\, V_{opt}(x,x')\exp(-i A_{0} x') F_{o}(x') \nonumber
\eeq
and it cannot simply be ``gauged'' away as in the BO approximation.
\item If the collision energy is sufficiently large so that $ E >> \Delta $ and both channels are
open the resulting scattering properties become independent of both $A_{0}$ and $A_{1}$. In the case where the
internal state energy defect $\Delta$ can be neglected, we can treat $ {\underline A}$ as a pure gauge, i.e. it
does not affect the scattering properties of the system.
\end{itemize}

\section{Geometric magnetism}
The  model introduced above has interesting and salient properties but it lacks the
feature of geometric magnetism.  Because the diagonal components $A_{0}$ can be
gauged away in the adiabatic, or BO, limit effective Lorentz forces do not manifest. Can we construct a gauge potential 
(again here we limit ourselves to the $h_{2}$ ) that
satisfies the integrability condition Eq. (\ref{1.5}), but includes diagonal components that are non-trivial? The
answer is yes but we need to extend our parameter space, at the least, into two dimensions i.e ${\bm R} = (x,y)$, $ -\infty < x,y < \infty$.
Consider 
\begin{widetext}
\beq
 U = 
\exp(-i \sigma_{3} \phi(x,y)) \exp(-i \sigma_{2} \Omega(x,y)) \exp(i \sigma_{3} \phi(x,y) )
\label{2.1}
\eeq
\end{widetext}
here $\sigma_{i}$ are the Pauli spin matrices and $\phi(x,y)$ and $\Omega(x,y)$ are arbitrary but single-valued
functions in the $xy$-plane.  

\subsection{Singular model}
By construction $U$ is single valued and the gauge potentials
derived from it satisfy integrability condition Eq. (\ref{1.5}). Using prescription (\ref{1.8}) we
find
\begin{widetext}
\beq {\bm { A}} = \left(
\begin{array}{cc}
 -2 \sin ^2(\Omega) {\bm \nabla \phi} & -\exp(-2 i \phi)  \left(\sin (2 \Omega ) {\bm \nabla} \phi +i {\bm \nabla \Omega} \right) \\
 \exp(2 i \phi)  \left(i {\bm \nabla} \Omega -\sin (2 \Omega) {\bm \nabla }\phi  \right) & 2 \sin
   ^2(\Omega) {\bm \nabla \phi}
\end{array}
\right).
\label{2.2}
\eeq  
\end{widetext} 
There is freedom of choice in $\Omega, \phi$, as long as single-valuedness is enforced,  but here let's introduce 
\beq
&& \phi(x,y)= L \, B_{0} \, y/2 \nonumber \\
&& \Omega(x,y) = \nonumber \\
&& \theta_{H}(x) \theta_{H} (L-x) Arcsin(\sqrt{x/L}) + \theta_{H} (x-L) \frac{\pi}{2}    
\label{2.3}
\eeq 
(A similar but not equivalent system was independently proposed in Ref.\cite{diaz11}) 
where $ \theta_{H} $ is the Heaviside step function and $ B_{0}, L $ are positive constants. Inserting Eq. (\ref{2.3}) into
Eq. (\ref{2.2}) we obtain,
For $ x < L $, ${\bm { A}} =0$, 
in the region $ 0 < x <L$,
\begin{widetext}
\beq
 {\bm { A}} = \left(
\begin{array}{cc}
 - x \, B_{0} {\bm {\hat j} } & -e^{-2 i \phi}  \left(\sin (2 \Omega ) \frac{L B_{0}}{2}  {\bm {\hat j}}  + 
 {\bf {\hat i}} \, \frac{i}{2 \sqrt{x(L-x)}} \right) \\
e^{2 i \phi}  \left(- \sin (2 \Omega ) \frac{L B_{0}}{2}  {\bm {\hat j}} 
 +  {\bf {\hat i}} \frac{ i }{2 \sqrt{x(L-x)}} \right)  & x \, B_{0} {\bm {\hat j} } 
\end{array}
\right) 
\label{2.4}
\eeq 
\end{widetext} 
and for $ x>L$ 
\beq
{\bm { A}} = {\bm {\hat j}} \left(
\begin{array}{cc}
 -  B_{0} L &  0 \\
0  & B_{0} L \end{array}
\right). 
\eeq
We keep $H_{BO}$ described in the previous paragraphs. In seeking scattering
solutions for this system it is again convenient to proceed in the adiabatic picture. The adiabatic
amplitudes obey equations (\ref{1.17}) except that the gauge potentials are now given by
Eq. (\ref{2.2}). As above, the ground internal
BO state is open and the excited state is closed and it is appropriate, for sufficiently low collision energies
and a large energy gap $\Delta$, to proceed with a BO projection of Eqs. (\ref{1.17}) into a single channel description.
Hence, for $ 0 < x < L$, we obtain
\beq
&& \partial^{2}_{x} F_{o} + (\partial_{y} - i {A}_{0})^{2} F_{o}  + k^{2} F_{o} - b(x) \, F_{o}=0 \nonumber \\
&& A_{0} =  B_{0} \, x   \nonumber \\
&& b(x)=  \sum_{k \neq i} {\bm A}_{ik} \cdot {\bm A}_{ki} =  \left [  \sin^{2}(2 \Omega) 
{\bm \nabla } \phi \cdot  {\bm \nabla } \phi + {\bm \nabla } \Omega \cdot {\bm \nabla } \Omega \right ]  = \nonumber \\
&& = x(L-x)  B_{0}^{2} + \frac{1}{4 x(L-x)}. \quad   
\label{2.5}
\eeq
This Schrodinger equation is equivalent to that obtained for a system in which a 
charged particle interacts with an effective magnetic field
\beq
{\bm B} = {\bm \nabla} \times \hbar {\bm A}_{0} = {\bm {\hat k} } \, \hbar B_{0}    \quad   0 < x  < L, 
\label{2.6}
\eeq
in addition to an induced scalar potential $b(x)$. The proposed spin 1/2 system does not have
an electromagnetic charge and so this effective magnetic field is a manifestation of what is, sometimes, called
artificial or synthetic magnetism\cite{juz06,lin09a,lin09b,simon11,spiel11,dal10b}.
 It leads to effective Lorentz type forces acting on the dynamical system.
The emergence of such forces was first demonstrated in molecular physics\cite{zyg87a,ber89,zyg90}.
\subsection{Non-singular model}
Whereas the above model describes quantum magnetism the induced scalar potential suffer singularities
at $x=0$ and at $x=L$ and so we wish to introduce a similar model that does not posses these singularities.
We now posit that
\beq
&& \phi(x,y)= L \, B_{0} \, y/2 \nonumber \\
&& \Omega(x,y) = \frac{\pi}{4} \Bigl ( 1 + \tanh[\beta x] \Bigr ).    
\label{3.1}
\eeq

We then obtain for the vector potential, $ {\bm A} = A_{x} \, {\bm {\hat i}} + A_{y} {\bm {\hat j}} $, where 
\begin{widetext}
\beq
A_{y}=\left(
\begin{array}{cc}
 -B_{0} L \sin^{2} \left(\frac{1}{4} \pi  (\tanh (\beta x)+1)\right) & -\frac{B_{0} L}{2} e^{-i
   B_{0} L y} \cos \left(\frac{1}{2} \pi  \tanh (\beta x)\right) \\
 -\frac{B_{0} L}{2} e^{i
   B_{0} L y} \cos \left(\frac{1}{2} \pi  \tanh (\beta x)\right)
 &
   B_{0} L \sin^{2} \left(\frac{1}{4} \pi  (\tanh (\beta x)+1)\right)
\end{array}
\right)
\label{3.2}
\eeq
\end{widetext}
and  $ A_{x}$ is
\beq
&& \left(
\begin{array}{cc}
 0 & -\frac{1}{4} i e^{-i B_{0} L y} \pi  \beta  \, sech^2(x \beta ) \\
 \frac{1}{4} i e^{i B_{0} L y} \pi  \beta  \, sech^2(x \beta ) & 0
\end{array}
\right). \nonumber \\
&& \label{3.3}
\eeq
In the limit $ x\rightarrow \infty$,
\beq
A_{y} \rightarrow 
\left(
\begin{array}{cc}
 -B_{0} L & 0 \\
 0 &
   B_{0} L 
\end{array}
\right),
\label{3.4} 
\eeq
whereas $ A_{x} \rightarrow 0$, and ${\bm A} \rightarrow 0$ as $ x \rightarrow -\infty $. 

\begin{figure}[ht]
\centering
\includegraphics[width=0.9\linewidth]{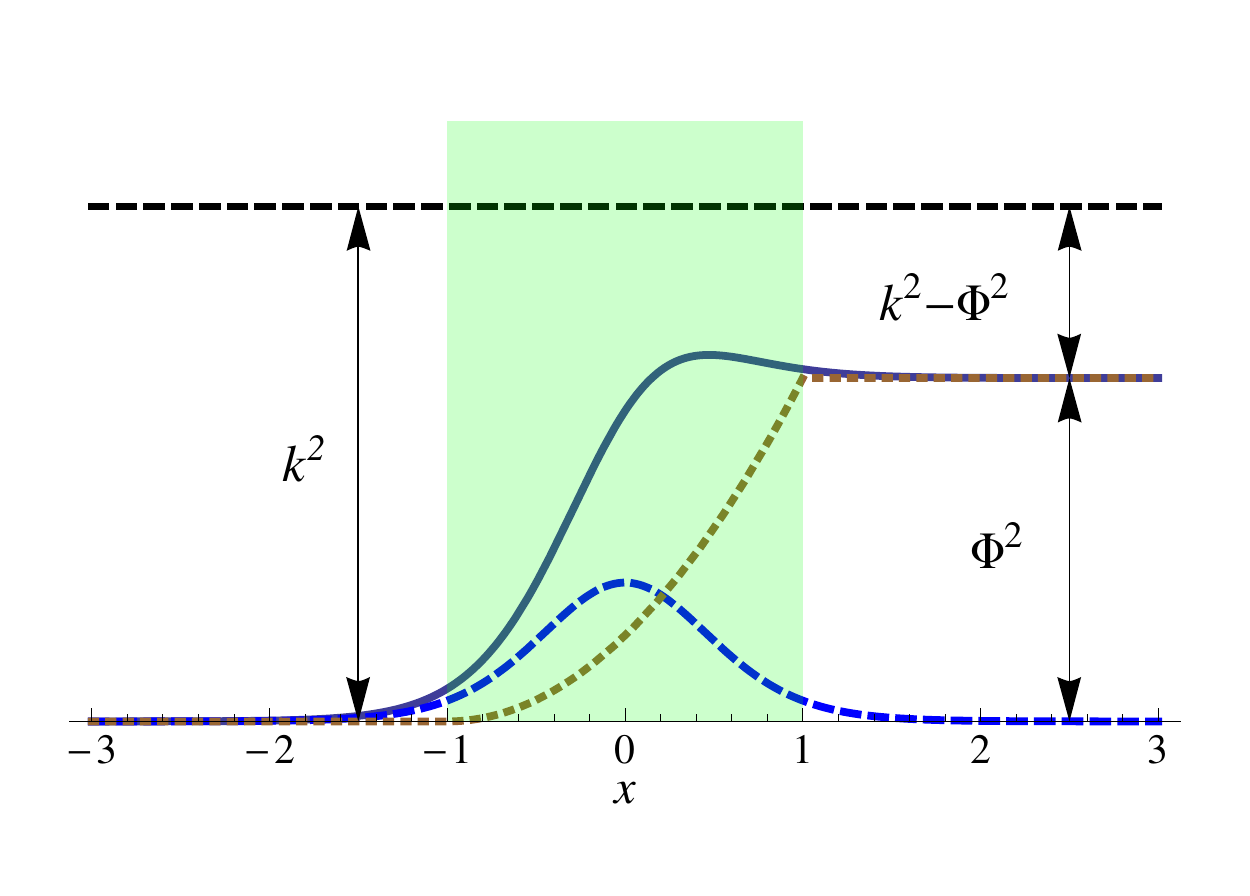}
\caption{\label{fig:fig1}(Color online) The solid blue line is a plot of Eq. (\ref{3.7}) the effective potential 
(vertical axis not shown) as a function of $x$ (horizontal axis). The dashed blue line represents
the effective scalar potential $b(x)$ given by Eq. (\ref{3.5b}). The dotted line represents the effective potential, given in Eq. (\ref{a.2}), for 
the ferromagnetic slab. The shaded region represent the slab.
We have chosen the parameters, $ \beta=1, L =2, B_{0}=1. $}
\end{figure}

The diagonal components of the ``magnetic'' induction $ |{\bm B}| \equiv \hbar \frac{\partial A_{y}}{\partial x}$ are
given by
\beq
B(x) = \mp \frac{\hbar}{4} \pi  \beta  B_{0} L \, sech^2(\beta  x) \cos \left(\frac{\pi}{2} \tanh (\beta  x) \right)  
\label{3.5}
\eeq
where the total magnetic flux density is
\beq
 \Phi \equiv \int_{-\infty}^{\infty} dx \, \frac{|B(x)|}{\hbar} = |B_{0}| L 
\label{3.5a}
\eeq
In the BO approximation, for the open channel, we obtain
\beq
&& \partial^{2}_{x} F_{o} + (\partial_{y} - i {A}_{0})^{2} F_{o}  + k^{2} F_{o} - b(x) \, F_{o}=0 \nonumber \\
&& A_{0}(x) =  \Phi \, \sin^{2} \left(\frac{1}{4} \pi  (\tanh (x \beta )+1)\right) \nonumber \\
&& b(x)= \sum_{k \neq i} {\bm A}_{ik} \cdot {\bm A}_{ki} =   \sin^{2}(2 \Omega) 
{\bm \nabla } \phi \cdot  {\bm \nabla } \phi + {\bm \nabla } \Omega \cdot {\bm \nabla } \Omega   = \nonumber \\
&&
\frac{1}{16} \left(2 \Phi^{2} (\cos (\pi  \tanh (\beta  x))+1)+\pi ^2 \beta ^2 \, sech^4(\beta 
   x)\right).
\label{3.5b}
\eeq
We assume the incoming wave is incident in the normal direction, i.e. $F_{o}=F_{o}(x)$ and the effective
equation for the BO amplitude is,
\beq
\partial^{2}_{x} F_{o} - {A}_{0}^{2}(x) \, F_{o}  + k^{2} F_{o} -  b(x) \, F_{o}=0, 
\label{3.6}
\eeq
or 
\beq
&& \partial^{2}_{x} F_{o} + k^{2}F_{0} - v_{eff}(x) F_{o} =0 \nonumber \\
&& v_{eff}(x) = {A}_{0}^{2}(x) +  \nonumber \\
&& \frac{1}{16} \left(2 \Phi^{2} (\cos (\pi  \tanh (\beta  x))+1)+\pi ^2 \beta ^2 \, sech^4(\beta 
   x)\right).  
\label{3.7}
\eeq
In Figure \ref{fig:fig1} we plot the effective potential for a given $\beta$ and $ \Phi \equiv B_{0} L$. 
We note that transmission occurs when $k^{2}$ is greater than the height of the potential
barrier $\Phi^{2} $ as $ x  \rightarrow \infty$. In that diagram we superimpose the effective potential discussed in the appendix
for the case of a ferromagnetic slab  characterized by the same value of $\Phi$. In the classical description the value
$ \Phi$ determines whether the particle has sufficient velocity so that its Larmor radius is greater than the slab thickness
and can thus penetrate the slab.  
Since $ {\bm A} \rightarrow 0$, as $ x\rightarrow -\infty$ we impose the incoming
boundary condition 
\beq
F_{0}(x) \rightarrow \exp(i k x) + r \exp(-i k x )
\label{3.10}
\eeq
where $r$ is the reflection coefficient. 
If $ k > \Phi$ transmission into the asymptotic region $ x \rightarrow \infty$ is allowed and, 
in this region, the vector potential has the form given by Eq. (\ref{3.4}).  
The outgoing asymptotic current is given by (see appendix A)
\beq
&& j_{x} =\frac{\hbar}{m} |t|^{2} \sqrt{k^{2}-\Phi^2} \nonumber \\
&& j_{y} = -\frac{\hbar}{m} A_{y} F_{o}^{*} F_{o} = -\frac{\hbar}{m} |t|^{2} \Phi
\label{3.8}
\eeq
where we made use, for $k > |\Phi|$, of the scattering boundary conditions
\beq
F_{o}(x)  \rightarrow  t \exp(i \sqrt{k^{2}-\Phi^{2}} x). 
\label{3.9}
\eeq  
The angle of deflection, with respect to the normal is then given by
\beq
|\tan\theta|=\frac{j_{y}}{j_{x}} = \frac{|\Phi|}{\sqrt{k^{2}-\Phi^{2}}}
\label{3.9a}
\eeq
and agrees with the result obtained in the classical description  of scattering of
a charged particle particle by a ferromagnetic slab (see Appendix B).
\subsection{Multichannel description}
\begin{figure*}[ht]
\centering
\includegraphics[width=0.7\linewidth]{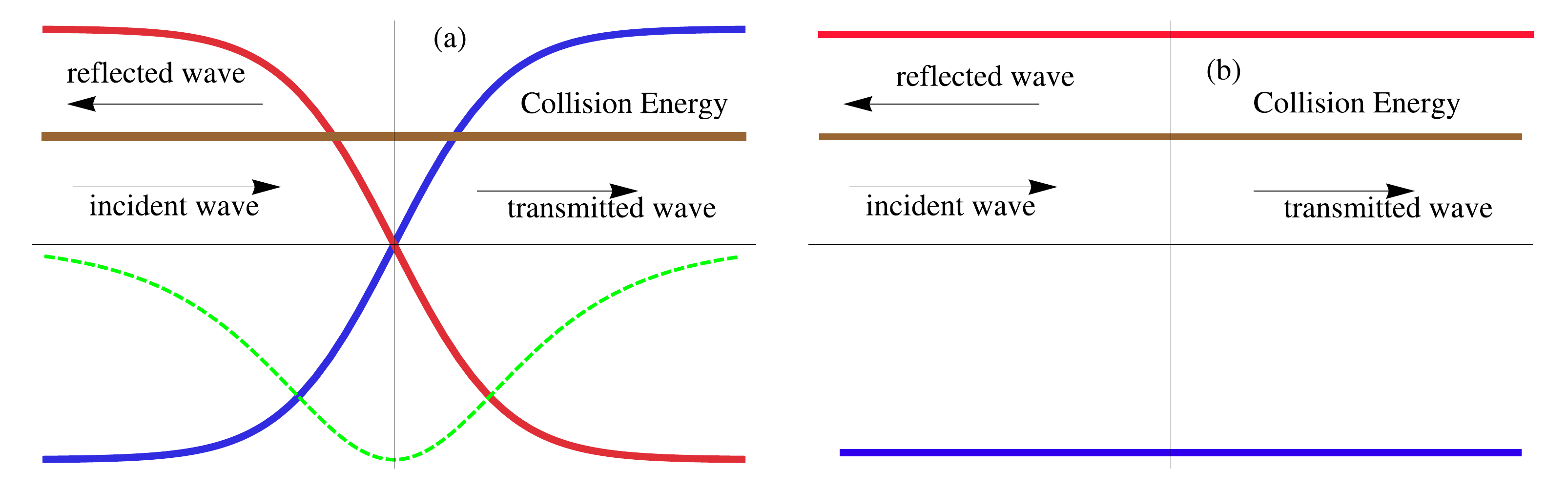}
\caption{\label{fig:fig2}(Color online) Schematic diagram of scattering setup described by Eqs. (\ref{4.2}). (a) An incident wave in the open channel whose potential energy is 
given by the solid blue line. The horizontal axis represent the $x$ coordinate, and for $ x<0$ the channel described by the red solid line
is closed. For $x>0$ the transmitted wave is propagated on the potential surface given by the red solid line. The brown line represent the total
collision energy of the system and the green dashed line represents the off-diagonal coupling between the two potential surfaces.
(b) The same system now illustrated in the adiabatic picture. The blue line is the BO energy for the open channel and the red line represents
the BO energy for the closed channel.}
\end{figure*}
The discussion presented above makes explicit use of the BO approximation, we now repeat the
analysis using the full, and exact, multichannel description. 
We need to solve the coupled equations,
\beq
&& -\frac{\hbar^{2}}{2 m} \Bigl ( {\bm \nabla} - i {\bm { A}} \Bigr )^{2} {\underline F} + { V} \, {\underline F} -E {\underline F}=0 \nonumber \\
&& {\underline F}= 
  \left ( \begin{array}{c}
 F_{c}\\
   F_{o} \end{array} \right ) \nonumber \\
&& { V}=\left ( \begin{array}{cc}
 \Delta & 0 \\
  0 &  -\Delta \end{array} \right ) 
\label{4.1}
\eeq
where $ {\bm { A}} $ is given by Eqs. (\ref{3.2}) and (\ref{3.3}). The diabatic gauge amplitude ${\underline G}$ 
is related to ${\underline F}$ via gauge transformation Eq. (\ref{2.1}) and it obeys 
\beq
&&-\frac{\hbar^{2}}{2 m} {\bm \nabla}^{2} {\underline G} + { W} \, {\underline G}   - E  {\underline G} =0 \nonumber \\
&& { W}= { U} \, { V} \, { U}^{\dag}= \nonumber \\
&& \left(
\begin{array}{cc}
 \Delta  \cos (2 \Omega (x)) & e^{-i \Phi y} \Delta  \sin (2 \Omega (x)) \\
 e^{i \Phi y} \Delta  \sin (2 \Omega (x)) & -\Delta  \cos (2 \Omega (x))
\end{array}
\right).
\label{4.2}
\eeq
In Figure \ref{fig:fig2}  we plot both the diagonal and off -diagonal components of the matrix-valued potential ${ W}$ using
the expression for $ \Omega(x)$  given by Eq. (\ref{3.1}) for $y=0$. 
It is somewhat more convenient to define a new amplitude so that $ {\underline G}= { {\tilde U}} \, {\underline G}'$ 
\beq
{ {\tilde U}} = \exp(-i \frac{\Phi \, y}{2} \sigma_{3}),
\eeq
where $\sigma_{3}$ is the diagonal Pauli matrix
and ${ W}$ is replaced
by 
\beq
&& { W} \rightarrow { W'} = { {\tilde U}}^{\dag} W { {\tilde U}} =\nonumber \\
&&  \left(
\begin{array}{cc}
 \Delta  \cos (2 \Omega (x)) & \Delta  \sin (2 \Omega (x)) \\
 \Delta  \sin (2 \Omega (x)) & -\Delta  \cos (2 \Omega (x))
\end{array}
\right)
\label{4.3}
\eeq
Thus
\beq
&&-\frac{\hbar^{2}}{2 m} \bigl ({\bm \nabla}-i {\tilde {\bm A}} \Bigr )^{2} {\underline G'} + { W'} {\underline G'}   - E  {\underline G'} =0 \nonumber \\ 
&& {\tilde {\bm A}} = i \, {\tilde U}^{\dag} {\bm \nabla} {\tilde U} = {\bm {\hat j}} \, \left(
\begin{array}{cc}
 \frac{\Phi}{2} & 0 \\
 0 & -\frac{\Phi}{2}
\end{array}
\right)
\label{4.4}
\eeq
or, letting ${\underline G}' = \left ( \begin{array}{c} f_{1} \\ f_{2} \end{array} \right ) $, 
\begin{widetext}
\beq
&& \frac{\partial^{2} f_{1}}{\partial x^{2}} + (\frac{\partial}{\partial y} - i \frac{\Phi}{2})^{2} f_{1} -\frac{2 m}{\hbar^{2}} \, \Delta  \cos (2 \Omega) f_{1} +
\frac{2 m}{\hbar^{2}} \, E f_{1} 
- \frac{2 m}{\hbar^{2}} \, \Delta  \sin (2 \Omega) f_{2}  =0  \nonumber \\
&& \frac{\partial^{2} f_{2}}{\partial x^{2}} + (\frac{\partial}{\partial y} + i \frac{\Phi}{2})^{2} f_{2} +\frac{2 m}{\hbar^{2}} \, \Delta  \cos (2 \Omega) f_{2} 
+\frac{2 m}{\hbar^{2}} \, E f_{2} 
- \frac{2 m}{\hbar^{2}} \, \Delta  \sin (2 \Omega) f_{1}  =0. 
\label{4.5}
\eeq
\end{widetext}
With the ansatz $ f_{1}(x,y) = \exp(-i \frac{\Phi}{2} y) f_{1}(x) $ and 
$ f_{2}(x,y) = \exp(-i \frac{\Phi}{2} y) f_{2}(x) $ we obtain,
\begin{widetext}
\beq
&& \frac{\partial^{2} f_{1}}{\partial x^{2}} - \Phi^{2}  f_{1} -\frac{2 m}{\hbar^{2}} \, \Delta  \cos (2 \Omega) f_{1} +\frac{2 m}{\hbar^{2}} \, E f_{1} 
- \frac{2 m}{\hbar^{2}} \, \Delta  \sin (2 \Omega) f_{2}  =0  \nonumber \\
&& \frac{\partial^{2} f_{2}}{\partial x^{2}}  +\frac{2 m}{\hbar^{2}} \, \Delta  \cos (2 \Omega) f_{2} +\frac{2 m}{\hbar^{2}} \, E f_{2} 
- \frac{2 m}{\hbar^{2}} \, \Delta  \sin (2 \Omega) f_{1}  =0. 
\label{4.5a}
\eeq
\end{widetext}
In the asymptotic limit $ x \rightarrow -\infty$ $ \Omega \rightarrow 0$, and Eqs. (\ref{4.5a}) reduce to 
\beq
&& \frac{\partial^{2} f_{1}}{\partial x^{2}} - \Phi^{2} f_{1} -\frac{2 m}{\hbar^{2}} \, \Delta f_{1} +\frac{2 m}{\hbar^{2}} \, E f_{1} =0  \nonumber \\
&& \frac{\partial^{2} f_{2}}{\partial x^{2}}  +\frac{2 m}{\hbar^{2}} \, \Delta f_{2} +\frac{2 m}{\hbar^{2}} \, E f_{2} =0 
\label{4.6}
\eeq
whose solutions are, in this limit,
\beq
&& {\underline G}'= \exp(-i \frac{\Phi}{2} y) \left ( \begin{array}{c}  r_{12} \, \exp(\kappa x) \\
                                           \exp(i k x) + r_{11} \exp(-i k x) \end{array} \right ) \nonumber \\
&& \kappa \equiv \sqrt{\frac{2m}{\hbar^{2}} (\Delta - E) + \Phi^{2}} \nonumber \\
&&  k \equiv \sqrt{\frac{2m}{\hbar^{2}} (E+\Delta)}   .
\label{4.7}
\eeq
We have chosen this boundary condition so that
\beq
&& {\underline G} = { \tilde{U} }{\underline G}' = \exp(-i \frac{\Phi \, y}{2} \sigma_{3}) {\underline G'}= \nonumber \\
&& \left ( \begin{array}{c}  \exp(-i \Phi y) \, r_{12} \, \exp(\kappa x) \\
                                           \exp(i k x) + r_{11} \exp(-i k x) \end{array} \right ) \label{4.8}
\eeq 
describes an incoming, normally incident, plane wave in the open channel.
In the region $x \rightarrow \infty$, $ \Omega \rightarrow \frac{\pi}{2}$, and Eq. (\ref{4.5}) reduces to
\beq
&& \frac{\partial^{2} f_{1}}{\partial x^{2}} - \Phi^{2} f_{1} +\frac{2 m}{\hbar^{2}} \, \Delta  f_{1} +\frac{2 m}{\hbar^{2}} \, E f_{1} =0  \nonumber \\
&& \frac{\partial^{2} f_{2}}{\partial x^{2}}  - \frac{2 m}{\hbar^{2}} \, \Delta  f_{2} +\frac{2 m}{\hbar^{2}} \, E f_{2} =0 
\label{4.9}
\eeq
whose solutions can be written in terms of scattering boundary conditions
\beq
&& {\underline G}'=\exp(-i \frac{\Phi}{2} y) \left ( \begin{array}{c}  t_{12} \, \exp(i k' x) \\
                             t_{11} \exp(-\kappa' x) \end{array} \right ) 
			     \nonumber \\
&& k' =\sqrt{k^{2}-\Phi^{2}} \quad \kappa'=\sqrt{\frac{2 m}{\hbar^{2}} (\Delta-E)}					    
\eeq
or, in this limit,
\beq
&& {\underline G}=  \exp(-i \frac{\Phi \, y}{2} \sigma_{3}) {\underline G}'= \nonumber \\
&& \left ( \begin{array}{c}  \exp(-i \Phi y) \, t_{12} \, \exp(i k' x) \\
                                           t_{11} \exp(-\kappa' x) \end{array} \right ).
\label{4.10}
\eeq
The conserved current is given by
\beq
{\bm j} = \frac{i \hbar}{2 m} \Bigl ( {\bm \nabla} {\underline G}^{\dag} \, {\underline G}  -  {\underline G}^{\dag} {\bm \nabla} \, {\underline G} \Bigr ).
\label{4.11}
\eeq
Inserting expression (\ref{4.8}) into Eq. (\ref{4.11}) we find that, in the asymptotic region $ x \rightarrow -\infty$, 
\beq
&& j_{x} = \frac{i \hbar}{2 m} \Bigl ( -2 i k+ 2 i k |r_{11}|^{2} \Bigr ) = \frac{\hbar k}{m} \Bigl ( 1-|r_{11}|^{2} \Bigr ) \nonumber \\
&& j_{y} = -\frac{\hbar}{2 m} |r_{12}|^{2} \exp(2 \kappa x)  \rightarrow 0.
\label{4.12}
\eeq
Like-wise in the region $ x \rightarrow \infty $ 
\beq
&& j_{x} = \frac{i \hbar}{2 m} \Bigl (-2 i k' |t_{12}|^{2} \Bigr ) = \frac{\hbar}{m} \sqrt{k^{2}-\Phi^{2}} \, |t_{12}|^{2}  \nonumber \\
&& j_{y} = \frac{i \hbar}{2 m} \Bigl ( 2 i \Phi |t_{12}|^{2} \Bigr )= -\frac{\hbar}{m} \Phi \, |t_{12}|^{2}. 
\label{4.13}
\eeq
We compare these solutions with those obtained in the BO approximation. According to Eq. (\ref{2.1}),
\beq
{\underline G}={ U } \, {\underline F}
\label{4.13a}
\eeq
or
\beq
&& \left ( \begin{array}{c} F_{c} \\ F_{o} \end{array} \right ) = \nonumber \\
&& \left ( \begin{array}{c}  \exp(-i \Phi \,  y) \Bigl ( f_{1}(x) \cos\Omega + f_{2}(x) \sin\Omega \Bigr ) \\
 -f_{1}(x) \sin\Omega + f_{2}(x) \cos\Omega \end{array} \right ).
\label{4.14}
\eeq
Carrying out these transformations, using Eq. (\ref{4.8}) and Eq. (\ref{4.10}), we find   
as $ x\rightarrow -\infty$
\beq
{\underline F}=\left ( \begin{array}{c} F_{c} \\ F_{o} \end{array} \right ) = 
\left ( \begin{array}{c}  \exp(-i \Phi y) \, r_{12} \, \exp(\kappa x) \\
                                           \exp(i k x) + r_{11} \exp(-i k x) \end{array} \right )
\label{4.15}
\eeq
and as $ x\rightarrow \infty$ 
\beq
\left ( \begin{array}{c} F_{c} \\ F_{o} \end{array} \right ) =\left(
\begin{array}{c}
 t_{11} \exp(-\kappa'  x) \exp(-i \Phi y) \\
 -t_{12} \exp(i \sqrt{k^{2}-\Phi^{2} } \, x)
\end{array}
\right).
\label{4.16}
\eeq
Projecting out the open channel amplitude (i.e. the Born-Oppenheimer amplitude) we obtain
\beq
&& F_{o} \rightarrow  \exp(i k x) + r_{11} \exp(-i k x)  \quad  x \rightarrow -\infty \nonumber \\
&& F_{o} \rightarrow  t_{12} \exp(i \sqrt{k^{2}-\Phi^{2} } x)  \quad  x \rightarrow \infty. 
\label{4.17}
\eeq
In Fig. (3) we present the result for the transmission coefficient 
\beq
|T| \equiv  \frac{|t_{12}|^{2} \sqrt{k^{2}-\Phi^{2}}}{k}
\label{4.18}
\eeq
and compare it to the expression obtained using the BO approximation where $t_{12}$ is replaced
by $t$ given in Eq. (\ref{3.9}). The numerical methodolgy, for producing the data shown in Fig. (3),
is included in the Supplementary Material enclosed.
The wavenumber $ k$ is expressed in terms of the flux density
$ \Phi $ and transmission is allowed for $k > \Phi$. In the coupled channel calculations 
we set the closed channel parameter $ \kappa' = k$, and so for larger $k$ the BO approximation
becomes valid in the limit as $ \kappa' \rightarrow \infty$. It is clear from the figure that the
predictions of the BO, or adiabatic, approximation are in very good agreement with the results obtained
using the fully coupled calculations. As $ \beta$ get large the transmission coefficient is
suppressed since the barrier height of the repulsive potential $b(x)$, given by Eq. (\ref{3.5b}), becomes larger. 

\begin{figure}[ht]
\centering
\includegraphics[width=1.0\linewidth]{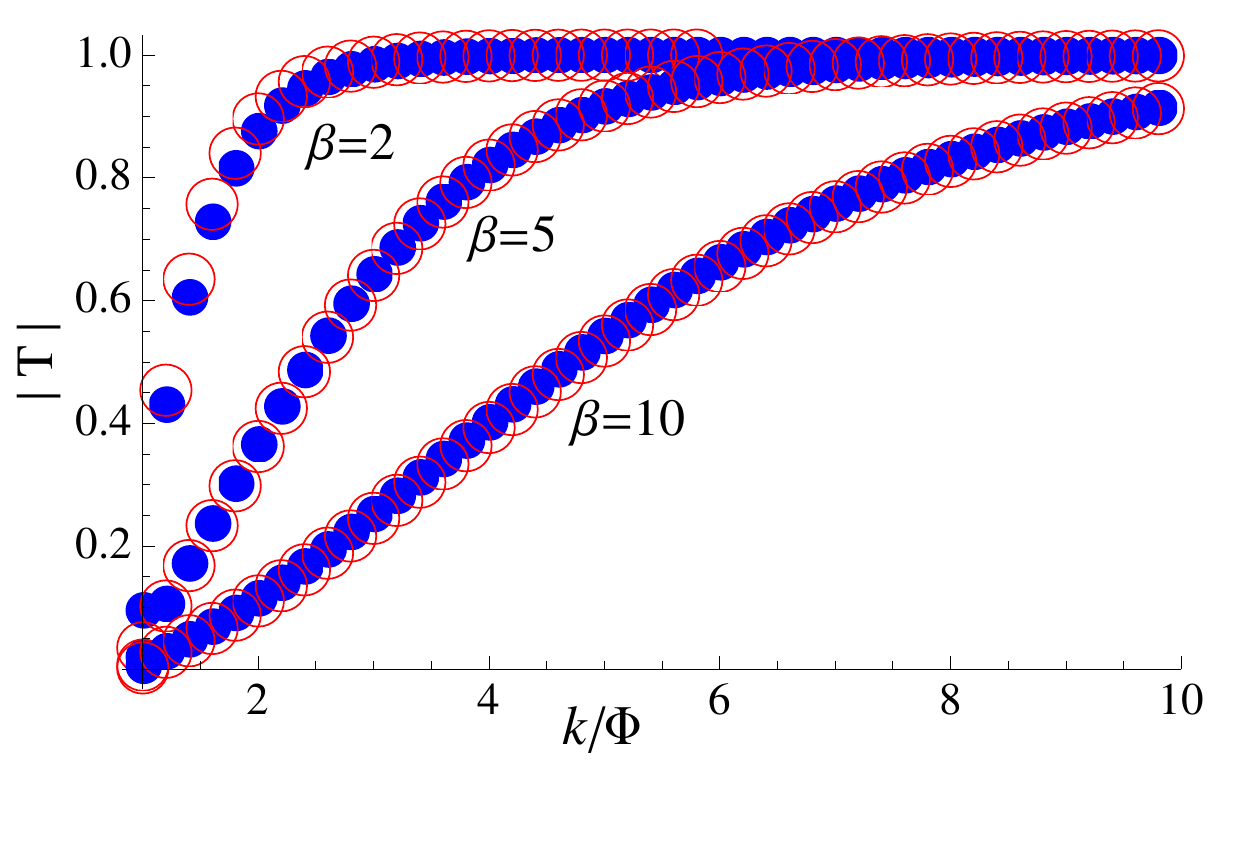}
\caption{\label{fig:fig3} (Color online) Transmission coefficient as a function of incident wavenumber $k$ expressed in units of $\Phi$ taken here to have the value of unity. The solid circles represent data obtained using the BO description, whereas the open circles
denotes the transmission coefficient obtained using the coupled channel theory and given by Eq. (\ref{4.18})}.
\end{figure}
We conclude that the result of the BO approximation, with the presence of a vector potential that describes
geometric magnetism, does indeed provide an accurate account for the dynamics and is in harmony with
the predictions of the exact treatment.

\subsection{Open channels}
\subsubsection{Coupled equations in the diabatic picture}
According to the above discussion the amplitudes in the diabatic gauge are given by
\beq
 {\underline G} = \left ( \begin{array}{c}  \exp(-i \Phi y) \, f_{1}(x) \\
                                          f_{2}(x) \end{array} \right )
\label{5.1}
\eeq
where $f_{1}(x),f_{2}(x)$ obey the coupled equations Eq. (\ref{4.5a}) and $\Phi \equiv B_{0} L$. The expression for the conserved
current is then given by
\beq
&& j_{x} =\frac{\hbar}{m} \Bigl ( Im[f_{1}^{*} \partial_{x} f_{1}] + Im[f_{2}^{*} \partial_{x} f_{2}] \Bigr ) \nonumber \\
&& j_{y} = - \, \frac{\hbar}{m} |f_{1}|^{2}. 
\label{5.2}
\eeq
We now evaluate the integral 
\beq
{\cal F} \equiv \int_{\cal{C}} \, d{\bm s} \cdot {\bm j} 
\label{5.3}
\eeq
for contour ${\cal C}$ shown in Figure \ref{fig:fig4}. In that figure the current streams are generated using the procedure
outlined in the supplementary material.
Consider the segments labeled I,III in the diagram. The contribution
to Eq. (\ref{5.3}) from these segments are
\beq
\frac{\hbar}{m} \int_{-w}^{w} dy \, \Phi \, \Bigl ( |f_{1}(a)|^{2} - |f_{1}(-a)|^{2} \Bigr ),
\label{5.4}
\eeq
like-wise for segment II we obtain
\beq
\frac{\hbar}{m} \int_{-a}^{a} dx \,  \Bigl ( Im[f_{1}^{*} \partial_{x} f_{1}] + Im[f_{2}^{*} \partial_{x} f_{2}] \Bigr ).   
\label{5.5}
\eeq
Since the integrand in Eq. (\ref{5.5}) is independent of $y$ this contribution is canceled by that
along segment IV. Therefore, as $ |a| \rightarrow \infty $  we find
\beq
\int_{C} \, d{\bm s} \cdot {\bm j} = \frac{\hbar}{m} \, 2 w \Phi \, |f_{1}(a)|^{2}=  \frac{\hbar}{m} \, 2 w \,\Phi \, |t_{12}|^{2}
\label{5.6}
\eeq
where we made use of Eq. (\ref{4.10}).
We now consider the case where $ E > \Delta + \frac{\hbar^{2} \Phi^{2}}{2 m} $ 
i.e. both channels in Eq. (\ref{4.5}) are open. As $ x \rightarrow -\infty$  Eq. (\ref{4.8}) is replaced
by 
\beq
{\underline G} \rightarrow \left (  \begin{array}{c}   \exp(-i \Phi y) \, r_{12} \, \exp(-i k_{2} x) \\
                                           \exp(i k x) + r_{11} \exp(-i k x) \end{array} \right )
\label{5.7}
\eeq
where 
\beq
k_{2} \equiv \sqrt{\frac{2 m}{\hbar^{2}}( E -\Delta)-\Phi^{2}}
\label{5.8}
\eeq
and which describes an incoming wave in channel 1 that can also
reflect into channel 2. Similarly, in the region $ x \rightarrow \infty$  we require
\beq
{\underline G} \rightarrow  \left (  \begin{array}{c}   \exp(-i \Phi  y) \, t_{12} \, \exp(i k' x) \\
                                           t_{11} \exp(i k'_{2} x) \end{array} \right )
\label{5.9}
\eeq
where
\beq
&& k' = \sqrt{k^{2}-\Phi^{2}} \nonumber \\
&& k'_{2}= \sqrt{\frac{2 m}{\hbar^{2}} (E-\Delta)}. 
\label{5.10}
\eeq					   
We  evaluate $ \int_{\cal{C}} \, d{\bm s} \cdot {\bm j} $ as above. Using Eqs. (\ref{5.2}) we find
 that the contribution from segments II,IV cancel and  that contributions
from segments I, III give
\beq
&& {\cal F}=\int_{\cal{C}} \, d{\bm s} \cdot {\bm j} = \frac{\hbar}{m} \Phi \, 2w \, \Bigl [ |f_{1}(a)|^{2} - |f_{1}(-a)|^{2} \Bigr ]
= \nonumber \\
&& \frac{\hbar}{m} \Phi \, 2w \, \Bigl [ |t_{12}|^2 - |r_{12}|^{2} \Bigr ].
\label{5.10a}
\eeq

\begin{figure}[ht]
\centering
\includegraphics[width=0.9\linewidth]{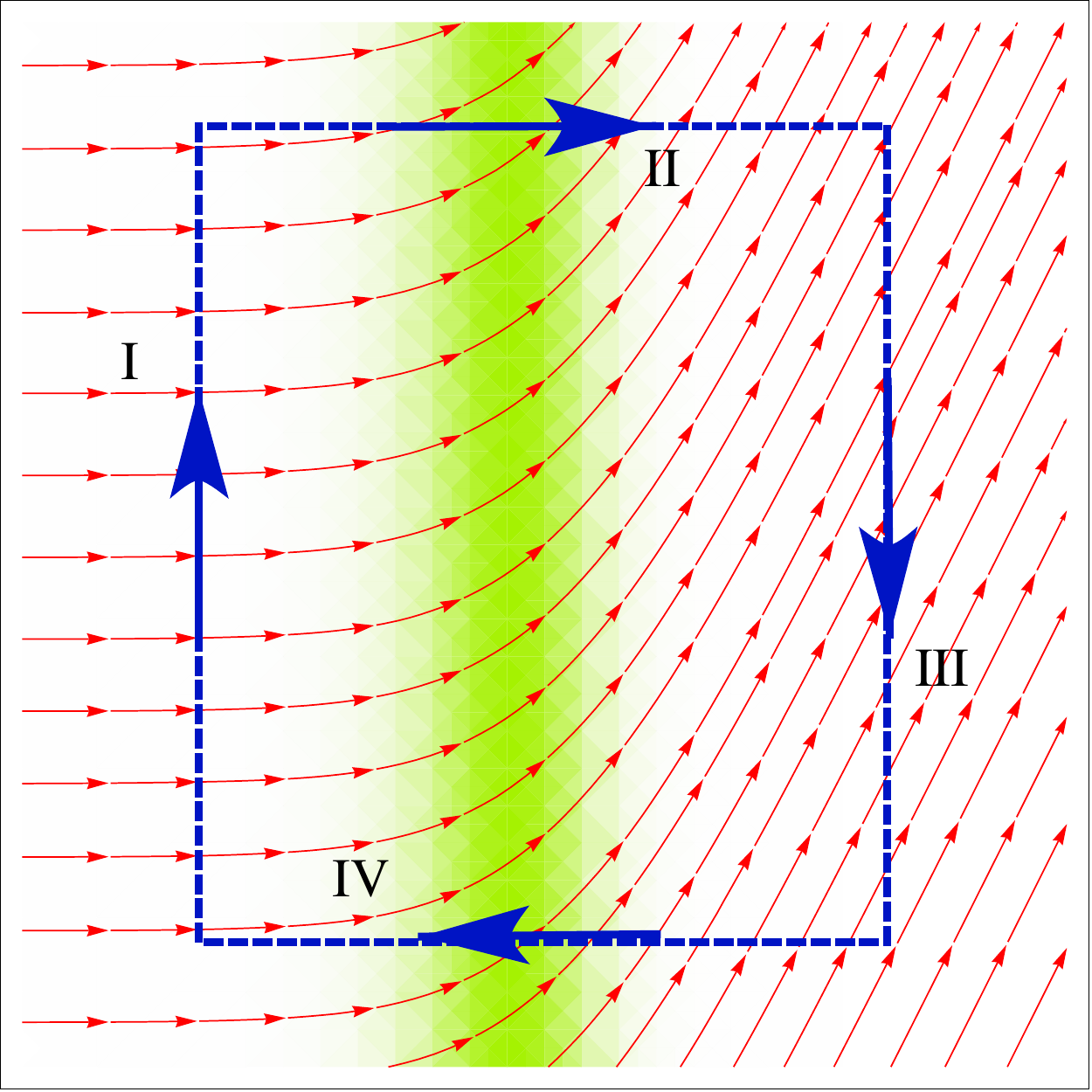}
\caption{\label{fig:fig4} (Color online) Contour plot of the probability current stream given by Eq. (\ref{4.11}).
The incident wave is along the horizontal.
 The shaded green
area represents region where the magnetic induction $B(x)$ given by Eq. (\ref{3.5}) is non-negligible.
Dashed blue line represents the contour ${\cal C}$. Segment I,III extend from $-w$ to $w$, and segments
II, IV range from $-a$ to $a$.
The collision energy is such that the upper adiabatic channel, 
shown in the diagram of Figure \ref{fig:fig2}, is closed. }
\end{figure}
To summarize, we found that
\beq
&& {\cal F} = \frac{\hbar}{m} \, 2 w \Phi \, |t_{12}|^{2} \quad  E < \Delta + \frac{\hbar^{2} \Phi^{2}}{2 m} \nonumber \\
&& {\cal F} = \frac{\hbar}{m} \, 2 w \Phi  \, \Bigl [ |t_{12}|^2 - |r_{12}|^{2} \Bigr ] 
\quad  E > \Delta + \frac{\hbar^{2} \Phi^{2}}{2 m}. 
\label{5.12}
\eeq
\subsubsection{Coupled equations in the adiabatic picture}
Before we discuss the significance of Eq. (\ref{5.12}), 
it is useful to re-derive this result in the adiabatic
representation. The gauge invariance of expression (\ref{5.3}) follows from general principles, but it is wortwhile
to investigate the role of the vector potential in its enforcement.
In the adiabatic gauge the amplitudes obey Eq. (\ref{4.1}) and lead to the conserved current
\beq
{\bm j} = \frac{i \hbar}{2 m} \Bigl ( {\bm \nabla} {\underline F}^{\dag} \, {\underline F}  -  {\underline F}^{\dag} {\bm \nabla} \, {\underline F} 
+2 i {\underline F}^{\dag}  {\bm {\underline A}} \, {\underline F} \Bigr )
\label{6.1}
\eeq
where ${\underline F}$ is given by Eq. (\ref{4.14}). Consider first
\beq
{\bm {\tilde j} } \equiv \frac{i \hbar}{2 m} \Bigl ( {\bm \nabla} {\underline F}^{\dag} \, {\underline F}  -  
{\underline F}^{\dag} {\bm \nabla} \, {\underline F}  \Bigr ).
\label{6.2}
\eeq
We show below that
\beq
&& \text{for} \quad  E > \Delta + \frac{\hbar^{2} \Phi^{2}}{2 m} \nonumber \\
&& {\tilde {\cal F}} \equiv  \int_{\cal{C}} \, d{\bm s} \cdot {\bm {\tilde j}} = 
\frac{\hbar}{m} \Phi \,2w \,\Bigl (|t_{11}|^{2} - |r_{12}|^{2} \Bigr )
 \nonumber \\
&& \text{and} \nonumber \\
&& {\tilde {\cal F}} = 0 \quad  E < \Delta + \frac{\hbar^{2} \Phi^{2}}{2 m}.
\label{6.3}
\eeq
The contribution to the left-hand side of Eq. (\ref{6.3}) from segment I is given by
\beq
&& \int_{-w}^{w} dy \, {\tilde j}_{y} = \frac{\hbar}{m} \int_{-w}^{w} dy (Im ({F_{o}}^{*} \partial_{y} F_{o}) + Im ({F_{c}}^{*} \partial_{y} F_{c}) )
= \nonumber \\
&& -\frac{\hbar}{m} \int_{-w}^{w} dy \, \Phi |f_{1}(-a)|^{2}  
\label{6.4}
\eeq
where we have used Eq. (\ref{4.14}) and the fact $\Omega \rightarrow 0 $ as $ a \rightarrow  -\infty$.
Like-wise, for segment III we obtain
\beq
-\int_{-w}^{w} dy \, {\tilde j}_{y} = \frac{\hbar}{m} \int_{-w}^{w} dy \, \Phi |f_{2}(a)|^{2}. 
\label{6.5}
\eeq
The contribution from segment II is 
\beq
&& \int_{-a}^{a} dx {\tilde j}_{x} = \nonumber \\
&& \frac{\hbar}{m} \int_{-a}^{a} dx (Im ({F_{o}}^{*} \partial_{x} F_{o}) + Im ({F_{c}}^{*} \partial_{x} F_{c}) ),
\label{6.6}
\eeq
but since it is independent of $y$ it cancels with the contribution arising from segment IV.
Inserting the limits for $ |f_{1}(-a)|^{2} $ and $ |f_{2}(a)|^{2} $, as $ |a| \rightarrow \infty$,
into Eq. (\ref{6.4}) and Eq. (\ref{6.5}) we arrive at our assertion Eq. (\ref{6.3}). 
We now evaluate
\begin{widetext}
\beq
&& {\cal F}_{A} \equiv  -\frac{\hbar}{m} \int_{\cal{C}} \, d{\bm s} \cdot {\underline F}^{\dag}  {\bm { A}} \, {\underline F} = 
-\frac{\hbar}{m} \int_{\cal{C}} \, d{\bm s} \cdot \Bigl [  {F_{c}}^{*} F_{c} {\bm A}_{11} + {F_{c}}^{*} F_{o} {\bm A}_{12}
+ {F_{o}}^{*} F_{c} {\bm A}_{21} + {F_{o}}^{*} F_{o} {\bm A}_{22} \Bigr ]
\label{6.7}
\eeq
\end{widetext}
where $ {\bm A}_{ij}$ are the matrix elements given in Eq. (\ref{2.2}). The contribution
from segment II,
\beq
&& i\frac{\hbar} {m}\int_{-a}^{a} dx \, \partial_{x} \Omega \, 
\Bigl ( {F_{c}}^{*} F_{o} \exp(-i \Phi w) - {F_{o}}^{*} F_{c} \exp(i \Phi w) \Bigr ) = \nonumber \\
&& i\frac{\hbar} {m}\int_{-a}^{a} dx \, \partial_{x} \Omega \, 
\Bigl ( f_{1}^{*}f_{2} - f_{2}^{*} f_{1}  \Bigr ),
\label{6.8}
\eeq
is independent of $y$ and it is
canceled by the contribution from segment IV. Consider the contributions
from segments I and III. First, we focus on terms that involve the off-diagonal elements,
\begin{widetext}
\beq
&& \frac{\hbar}{m} \frac{\Phi}{2}  \int_{-w}^{w} dy \sin(2 \Omega) 
\Bigl ( {F_{c}}^{*} F_{o} \exp(-i \Phi y) + {F_{o}}^{*} F_{c} \exp(i \Phi y) \Bigr ).
\label{6.9}
\eeq
\end{widetext}
For segments I,III, $ \Omega=0,\frac{\pi}{2}$ respectively and according to Eq.(\ref{6.9}) these
contributions vanish. Finally, we consider the diagonal terms,
\beq
\frac{\hbar}{m} \Phi \int_{-w}^{w} dy \sin^{2}(\Omega) \Bigl ( {F_{c}}^{*} F_{c} - {F_{o}}^{*} F_{o} \Bigr )
\label{6.10}
\eeq
whose non-vanishing contribution comes from segment III and has
the value given by 
\beq
{\cal{F}}_{A} = \frac{\hbar}{m} \Phi \int_{-w}^{w} dy  \Bigl ( |f_{1}(a)|^{2}  - |f_{2}(a)|^{2} \Bigr ).
\label{6.11}
\eeq
Using Eqs. (\ref{4.14}) and (\ref{5.1}) we obtain
\beq 
{\cal F}_{A}= && \frac{\hbar}{m} \Phi \, 2w \, \Bigl ( |t_{12}|^{2} - |t_{11}|^{2} \Bigr )  \quad  E > \Delta + \frac{\hbar^{2} \Phi^{2}}{2m} \nonumber \\
{\cal F}_{A}=  && \frac{\hbar}{m} \Phi \, 2w \,  |t_{12}|^{2}   \quad  E < \Delta + \frac{\hbar^{2}\Phi^{2}}{2m}. 
\label{6.11a}
\eeq
Adding this contribution to that given by Eq. (\ref{6.3}) we obtain,
\beq
&& {\cal F}={\tilde {\cal F}} + {{\cal F}}_{A} = \frac{\hbar}{m} \Phi \, 2w \,
\Bigl (|t_{12}|^{2} - |r_{12}|^{2} \Bigr )  \quad  E > \Delta + \frac{\hbar^{2}\Phi^{2}}{2m} \nonumber \\
&& {\cal F} =  \frac{\hbar}{m} \Phi \, 2w \, |t_{12}|^{2} \quad E < \Delta + \frac{\hbar^{2}\Phi^{2}}{2m}
\label{6.12}
\eeq
and which agrees with expression (\ref{5.12}) obtained in the diabatic picture.

Equation (\ref{6.12}) constitutes one of the main results in this paper. It
expresses the scattering solutions in terms of the integral
\beq
{\cal F} \equiv \int_{\cal C} d {\bm s} \cdot {\bm j} 
\label{6.13}
\eeq
which is invariant with respect to the gauge or representation. That is, 
${\cal F}$ has the same value in both the diabatic or adiabatic pictures.
If the collision energy is such that the excited adiabatic channel is closed, then
$ {\cal F} $ does not vanish and its value is given by the second line in Eq. (\ref{6.12}). Furthermore,
in the adiabatic representation that value is determined from the contribution of the
current that is proportional to $ {\bm A}$. Because this result is obtained
in the, exact, coupled-channel framework it does not require the invocation of the BO approximation.
However, the BO approximation provides an elegant and accurate description and
the appearance of a vector gauge potential in Eq. (\ref{3.5b}) is simply a consequence of the fact
that the BO approximation is faithful to the requirement that ${\cal F}$ does not have a null value. 
Indeed, the arbitrary removal
of the vector potential in the BO approximation compromises its efficacy.

It is important to note that the vector potential Eq. (\ref{3.5b}) and the resulting
effective ${\bm B}$ field are source-less and there are no-singularities associated with it.
Consider now the case where the BO approximation is no longer valid at higher collision energies
where the excited adiabatic state is also open. In the high energy limit where $ E>> \Delta$ 
we can ignore $\Delta$ (in lowest order) and it is evident that in this limit
\beq
&& |t_{12}| \rightarrow |r_{12}| \rightarrow 0 
\label{6.14}
\eeq
and $ {\cal F} \rightarrow 0$. Thus, at higher collision energies
the scattering properties are not affected by the presence of the vector
gauge potential, i.e. its influence on the system is what one would expect from
that of a pure gauge. This does not preclude the fact that when working in the
adiabatic picture one cannot ignore the gauge potentials when solving for the multi-channel amplitudes.
An interesting case concerns the energy regime
where both channels are open, thus voiding the BO approximation,
but the limit behavior, given in Eq. (\ref{6.14}), has not been reached.
 That discussion will be deferred to future investigations by the author.

In summary we have shown the necessity of including the non-trivial vector potential Eq. (\ref{3.5b}) in
the BO approximation. The potential predicts a Lorentz force that determines the scattering properties of the proposed system.
We illustrated how this system is an analog to that of a charged particle
that is scattered by a ferromagnetic slab. We have solved the, exact, fully coupled equations in both the diabatic and
adiabatic pictures and showed the fidelity of the BO approximation at low collision energies. At higher collision
energies we find, in an appropriate limit, that the presence of the induced gauge potentials do not affect 
scattering properties. Thus the emergence of non-trivial gauge structure in the low energy limit
is suggestive of phenomena that arise when a symmetry is broken\cite{nambu61}.     

\subsection{Time-dependent treatment}
In the previous section we investigated the scattering solutions of Eq. (\ref{4.1}) using time independent methods.
In this section we exploit  time-dependent methods to enhance and generalize the conclusions of the previous sections.
We consider the time-dependent version of coupled equations (\ref{4.2}) for the amplitude $\psi$,
\beq
&& i \hbar \, \frac{\partial f}{\partial t} = -\frac{\hbar^{2}}{2 m} {\bm \nabla}^{2} f + V \, f + V_{12} g \nonumber \\
&& i \hbar \, \frac{\partial g}{\partial t}= -\frac{\hbar^{2}}{2 m} {\bm \nabla}^{2} g + V^{*}_{12} f - V \, g
\label{O1.4}
\eeq
where 
\beq
&& \psi \equiv \left ( \begin{array}{c} f \\ g \end{array} \right ) \nonumber \\
&& V= \Delta \, \cos(2 \Omega(x)) \nonumber \\
&& V_{12}= \exp(- \, i \, \Phi \, y) \, \Delta \, \sin(2 \Omega(x)). 
\label{O1.5}
\eeq
$\Omega(x)$ is defined in Eq. (\ref{3.1})
and $ \Phi \equiv B_{0} L$. Eqs. (\ref{O1.4}) may be solved numerically using a procedure
introduced by Hermann and Fleck \cite{fleck88} and discussed here in the Supplementary material.
In Figure (\ref{fig:figO2}a) we provide a time series contour plot of the probability densities $|f|^{2}$ and $|g|^{2}$.
At $\tau=0$ we place a Gaussian wave-packet $\psi_{0}=\left ( \begin{array}{c} 0 \\ g_{0} \end{array} \right )$  centered 
$\xi=-4.0, \, \eta=0$, where $(\xi,\eta)=(x/L,y/L)$ with $L$ an arbitrary length scale are dimensionless,
with an initial velocity directed along
the positive $\xi$ axis. In the region ($ \Omega \rightarrow 0 $)
\beq
 H_{ad} = \left ( \begin{array}{cc} \Delta & 0 \\   0 & -\Delta  \end{array} \right ) 
\nonumber
\eeq
and the wave packet evolves as that of a free particle until it reaches the interaction region  $ \xi \approx 0$. 
The wave-packet is illustrated by the blue contours in Figure (\ref{fig:figO2}a).
The initial kinetic energy of the packet was chosen so penetration of the potential barrier, illustrated in Figure (\ref{fig:fig2}), 
is prevented. However,  the packet can execute a transition into the open  channel across the barrier. In other words, a transition
from the $ g $ to $ f$ channel  occurs in the region $\xi \approx 0$. This is illustrated in Figure (\ref{fig:figO2}a) by the red
contours that represent the wave-packet, probability, contours in the $f$ channel. In addition to distortion and spreading of the packet there
is a noticeable swerve in its velocity as it emerges from the interaction region.
\begin{figure*}[ht]
\centering
\includegraphics[width=0.5\textwidth]{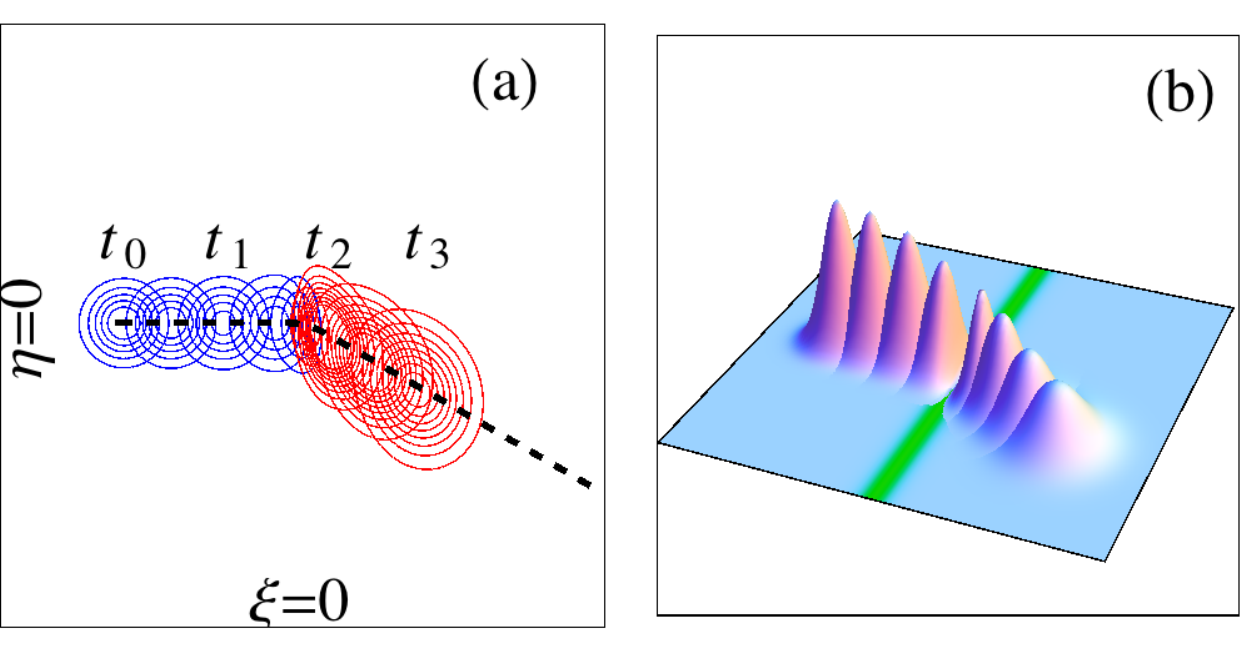}
\caption{\label{fig:figO2} (color online) (a) Circular (blue) contours represent the initial $g$ component of the wave packet probability distribution.
 At time $ t_{2}$ the packet executes a transition into the $f$ channel, shown by the elongated (red) contours, which is energetically open in the region $ \xi >0$. 
Subsequent to the transition the packet evolves as a free particle but with a pronounced swerve in its velocity. 
(b) 3D plot of the adiabatic gauge probability distribution  
$|\tilde g(t)|^{2}$. In both figures the deflection angle has the value $\tan \theta \approx 0.59 $. 
 In these calculations we took $k=12,\Phi=6,\Delta=200,\beta=2$ and $ {\tau}_{0}=0, {\tau}_{1} =0.1,{\tau}_{2}=0.15,{\tau}_{3}=0.3$. $\xi,\eta$ range between $\mp 2\pi$.   }
\end{figure*}
We define the adiabatic amplitudes,
\beq
{\tilde f} = f \, \cos\Omega(\xi) + \exp(-i \Phi \eta) \sin\Omega(\xi) g \nonumber \\
{\tilde g} = g \, \cos\Omega(\xi) - \exp(i \Phi \eta) \sin\Omega(\xi) f
\label{O1.7}
\eeq
and
${\tilde \psi} = \left ( \begin{array}{c} {\tilde f} \\ {\tilde g} \end{array} \right ) $ obeys the time-dependent
analogue of Eq. (\ref{4.1})
\beq
i \hbar \frac{\partial {\tilde \psi }}{\partial t} =
-\frac{\hbar^{2}}{2 m} \Bigl ( {\bm \nabla }- i {\bm A}  \Bigr )^{2} {\tilde \psi} + H_{BO} {\tilde \psi},   
\label{O1.8}
\eeq
where $ {\bm A}$ is a non-Abelian, pure, gauge potential. In the region $ \xi \rightarrow -\infty$, $ \Omega \rightarrow 0$
and ${\tilde g}  \rightarrow g$. Likewise as $ \xi \rightarrow \infty$, $ \Omega \rightarrow \pi/2 $ and 
$ {\tilde g} \rightarrow f $. This behavior is illustrated in Figure (\ref{fig:figO2}b) where we present a 3D plot
for the evolution of $ |{\tilde g}|^{2}$. In the adiabatic picture the open channel amplitude ${\tilde g}$ evolves
in a  constant adiabatic potential $ -\Delta $ shown in Figure (\ref{fig:fig2}b).  As long as the collision energy is below
the threshold for excitation into the upper adiabatic, or closed, channel the system evolves on a single adiabatic
surface. Under such conditions the Born-Oppenheimer (BO) approximation to the solutions of Eq. (\ref{O1.8}) is
appropriate. In this approximation, the projection operator $ P {\tilde \psi} = {\tilde g}$, is applied on 
 Eq. (\ref{O1.8}) to get
 \beq
 i \hbar \frac{\partial {\tilde g }}{\partial t} = -\frac{\hbar^{2}}{2 m}  \Bigl ( {\bm \nabla }- 
i {\tilde {\bm  A} }  \Bigr )^{2} {\tilde g} -\Bigl  (\Delta- \frac{\hbar^{2} \, b(x)}{2 m}  \Bigr )  {\tilde g}  
\label{O1.9}
\eeq
where $  {\tilde {\bm A} } = P \,  {\bm A} \, P  $ is an Abelian gauge potential with non-vanishing curl and $b(x)$
 is the induced scalar potential in Eq. (\ref{3.5b}). 
It leads to an effective magnetic induction 
given by Eq. (\ref{3.5}) 
and mimics
that incurred on a charged particle that is scattered by a ferromagnetic medium.
 The magnetic induction is normal to the plane of the page, 
 and is illustrated by the green shaded area in Figure (\ref{fig:figO2}b).
In Figure (\ref{fig:figO2}a) we also plot, shown by the dashed line, the 
trajectory for the solution of the classical equations of motion, subjected to a Lorentz force $ {\bm v} \times {\bm B}$, where
$|{\bm B}|$ is given by Eq. (\ref{3.5}).
Comparison of the classical path and that traced by 
the centers of the wave-packets shows good agreement.
\begin{figure}[ht]
\centering
\includegraphics[width=0.9\linewidth]{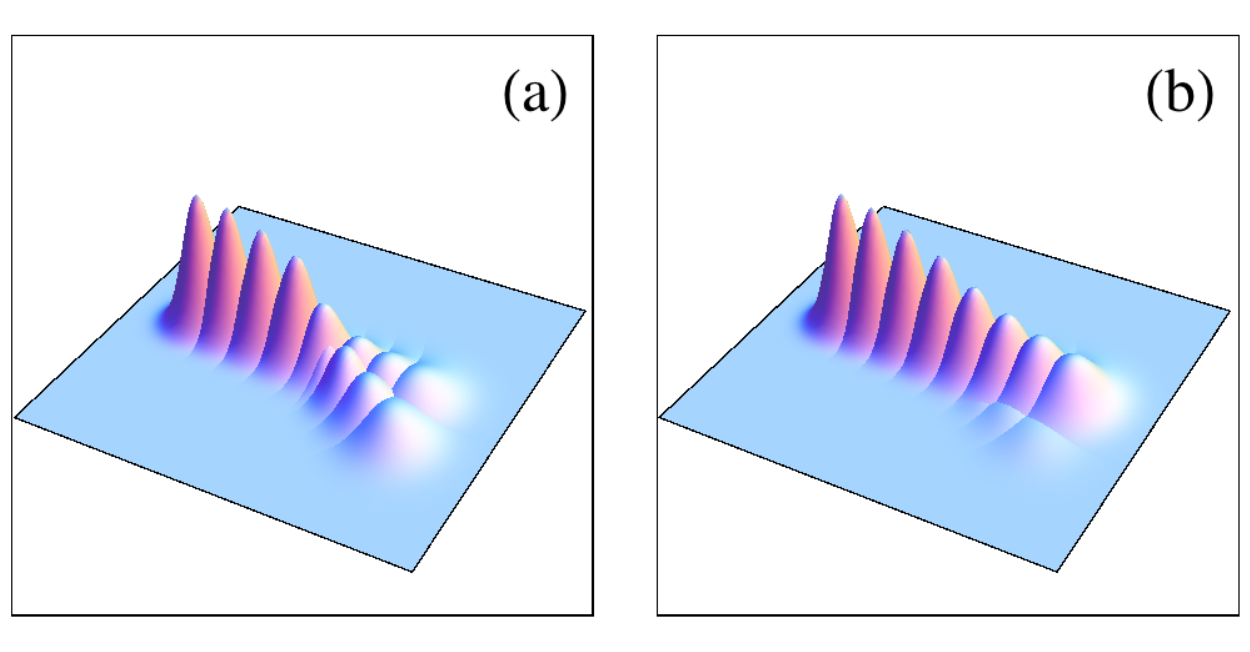}
\caption{\label{fig:figO3} (color online) Plot of probability densities $|g|^{2},|f|^{2}$ when both channels are open. The threshold energy $k_{t}^{2}=2 \Delta$. 
(a) $ k^{2} = 3 k_{t}^{2}$. (b) $k^{2}= 6 k_{t}^{2}$.} 
\end{figure}
The deflection angle 
suffered by a charged particle that is normally incident on a ferromagnetic slab, of finite width, 
with constant magnetic induction ${\bm B}$ directed
perpendicular to the plane of the page is given by Eq. (\ref{3.9a}).

In Table \ref{tab:Table1} we tabulate values of the deflection angles $\theta$, obtained by calculating the expectation values
$<\xi(t)>$, $<\eta(t)>$ for various values of incident, adiabatic, packet wave numbers
$ k$ and $\Phi$. In that table we show the dependence of $\theta$ on the choice of the energy gap parameter $\Delta$.
At lower collision energies, so that $ k^{2} << 2 \Delta $, we find that Eq. (\ref{3.9a})  provides a good approximation for $\theta$.
As the energy gap $2 \Delta$ is decreased, for a fixed value of $k$, Eq. (\ref{3.9a}) is less accurate.
However, even at threshold $ k^{2} \approx 2 \Delta$ there is still fairly good agreement between the calculated value and that predicted
by solutions of Eq. (\ref{O1.9}). When $ k^{2} > 2 \Delta$ the excited adiabatic state
is open and transitions from the adiabatic channel labeled $\tilde g$ into $\tilde f$ is energetically allowed. In Figure (\ref{fig:figO3}a) we
illustrate the evolution of the amplitudes $ |g(t)|^{2} $ and $ |f(t)|^{2} $ for the collision energy where $ 2 \Delta/k^{2} = 1/3 $. The
incident packet, in the $g$ channel, bifurcates when it reaches the interaction region. Because there is sufficient kinetic energy, the
remainder of the initial packet proceeds along the path $\eta=0$ in the region $ \xi >0$. However, a fraction of that packet makes
a transition into the $f$ channel, and our calculations show that the angle of the swerve illustrated in that figure is in harmony
with that obtained at the lower collision energies tabulated in Table \ref{tab:Table1}. Therefore there is a state-dependent spatial
segregation of the initial beam, a hallmark of quantum control. In panel (b) of that figure we plot
these probabilities for energies $ 2 \Delta/k^{2} = 1/6$ and now find a small, barely noticeable, remnant of the packet in the $f$ channel.
In the limit $ k \rightarrow \infty$  (or $\Delta \rightarrow 0$) Eq. (\ref{O1.4}) allow analytic solutions and $g(t)$ simply
evolves as that of a free particle. According to definition Eq. (\ref{O1.7} ) the initial, adiabatic gauge, packet $ {\tilde g}(t) $ makes 
a transition into the $ {\tilde f}$ channel in the region $\xi > 0$. This ''transition'' is induced by the off-diagonal
gauge couplings in Eq. (\ref{O1.8}). The  "transition" is simply an artifact of the adiabatic gauge (i.e. different definitions
for the scattering basis in the two asymptotic regions, $\xi<0$, $\xi>0$). 
\begin{table}
\caption{\label{tab:Table1} Calculated values for the deflection angle $\theta$ are tabulated
in the third column. The fourth column gives the values obtained using Eq. (\ref{3.9a})}.
\begin{ruledtabular}
\begin{tabular}{llll}
$2 \Delta/k^{2} $  &  $\Phi/k $  &  $\tan \theta $ &  $\tan \theta_{c} $   \\
\hline
$25/9$ & $1/2$ & $0.587$  &  $0.577  $    \\
$ 25/9$ & $1/4$ & $0.270$ & $0.258 $ \\ 
$ 25/9$ &  $1/12$ & $0.088  $ & $ 0.084 $ \\
\hline
$1$ & $1/2$ & $0.63$  &  $0.577  $    \\
$1$ & $1/4$ & $0.269$ & $0.258 $ \\ 
$1$ &  $1/12$ & $0.088  $ & $ 0.084 $ \\
\end{tabular}
\end{ruledtabular}
\end{table}
 Because of relation (\ref{1.3}) we conclude that ${\bm A}$ is
encoded in the definition of $H_{ad}$ and since $ [H_{BO},{\bm A}] \neq 0$ gauge symmetry is explicitly broken by $H_{BO}$.
Though ${\bm A}$ is trivial, in the sense of it being a pure gauge, 
quantum evolution selects and is sensitive to the projected ${\tilde {\bm A}} = P {\bm A} P$ {\bf \it non-trivial} connection.  In the
adiabatic picture the gauge potentials are explicit, being minimally coupled to the amplitudes.  As $ k \rightarrow \infty$, or $ \Delta \rightarrow 0$,
their presence simply contributes to a multichannel, or non-Abelian, phase in the adiabatic amplitudes that has
 no physical import. In contrast,
at lower energies the system behaves as if it has acquired a non-integrable phase factor. The effect is most pronounced when the excited
adiabatic state is closed.
 
\begin{figure}[ht]
\centering
\includegraphics[width=0.9\linewidth]{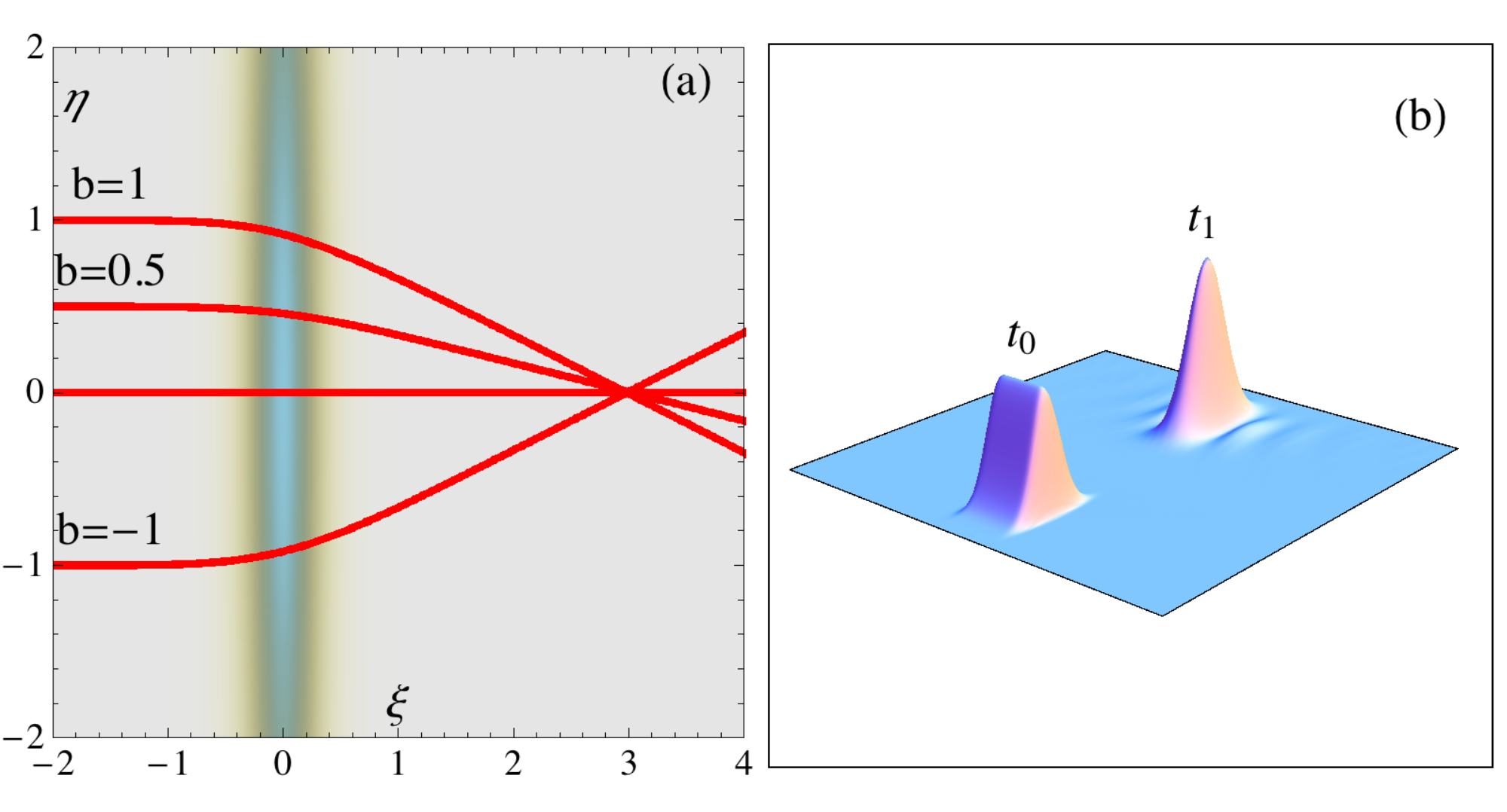}
\caption{\label{fig:figO4} (color online)  (a) Trajectories of wave-packet expectation values for various values of impact parameter. The shaded region
is a density plot of Eq. (\ref{O2.2}) for the ``magnetic'' induction. The horizontal axis gives the $\xi$ coordinate and the $\eta$ coordinate
is represented by the vertical axis.
(b) A wave packet slab having width $d=2$ along the $\eta$ axis at ${\tau}_{0}=0$
is propagated to the position shown at ${\tau}_{1}=0.3$.}
\end{figure}

 The time-dependent approach can be exploited for more complex scattering scenarios, for 
 consider the following form of the parameter 
\beq
\Phi(\eta) = \frac{\eta \, k}{\sqrt{\eta^{2} + 4 \gamma f^{2}}}
\label{O2.1}
\eeq
where $\gamma$ is, in general, a complicated function of $\beta, k, \Delta$. Here 
we set it to have the constant  value $\gamma =1$. Using Eq. (\ref{O2.1}) we propagate wave packets for various
values of impact parameter. In Figure (\ref{fig:figO4} a) we plot trajectories of the total expectation values
$ <\psi(t)|\xi|\psi(t)>, <\psi(t)|\eta|\psi(t)>$ for the various impact parameters b. At each impact
parameter we choose identical wave-packet widths and set $k=12, \Delta=400$. The trajectories shown
in that figure, by the solid red lines, demonstrate that the paths converge to a common focal point given by $f=3$. 
This result is gauge invariant, i.e. it can
be obtained using amplitudes obtained in either diabatic or adiabatic gauges. However, 
the adiabatic picture provides a transparent physical description. For, in it, the system is accurately described by
Eq. (\ref{O1.9}). That description includes the emergence of an effective magnetic induction ${\bm B}={\nabla} \times {\tilde {\bm A}}$ whose
magnitude is
\beq
\frac{\pi  \beta  k \eta \left(8 f^2+\eta^2\right) \text{sech}^2(\beta  \xi) \cos
   \left(\frac{1}{2} \pi  \tanh (\beta  \xi )\right)}{4 \left(4 f^2+\eta^2\right)^{3/2}}
\label{O2.2}
\eeq
and is normal to the plane of the page. In Figure (\ref{fig:figO4}b) we illustrate the propagation of a coherent wave packet slab of finite width
along the $\eta$ direction. After its passage through the ``magnetic'' lens at $\xi \approx 0$,  its shape is significantly distorted.
 At $\tau_{1}=0.3$, where the packet describes free particle evolution, it assumes the shape of a ``shark-fin'' as
shown in that figure. The width, along the $\eta$ direction, is significantly reduced from its value at $\tau_{0}$. A dramatic consequence
of the proposed ``magnetic'' lensing effect. Such a lens, if realized, could find application as an ``optical'' component in an atom laser.
In addition, consider two localized but coherent packets spatially separated at $t=0$.  After passing through the lens they meet and interfere.
Because of different geometric phase histories the interference pattern depends on the ``magnetic'' flux enclosed
by the paths. One can therefore anticipate its application as a novel expression
of atom interferometry. 
\subsection{Aharonov-Bohm gauge potentials}
The systems in the previous sections provides illustrations
of how effective gauge forces manifest in simple quantum systems, and serves as a template
for more intricate systems.
In the examples discussed above, we took for the internal Hamiltonian Eq. (\ref{1.2}) where $U$ is given
by Eq. (\ref{2.1}). Such a Hamiltonian may be engineered by ``dressing'' a 2-level atom with a laser\cite{dal10b,diaz11}.
Alternatively, one can dress a spin-1/2 system that has a magnetic moment with an external magnetic field,
as is the case with Berry's Hamiltonian\cite{ber84}. Below we describe a 2D version of such systems. A similar, field-theoretic, analog for such a system was also proposed in Ref.\cite{russel}.

\begin{figure}[h!]
\centering
\includegraphics[width=0.9\linewidth]{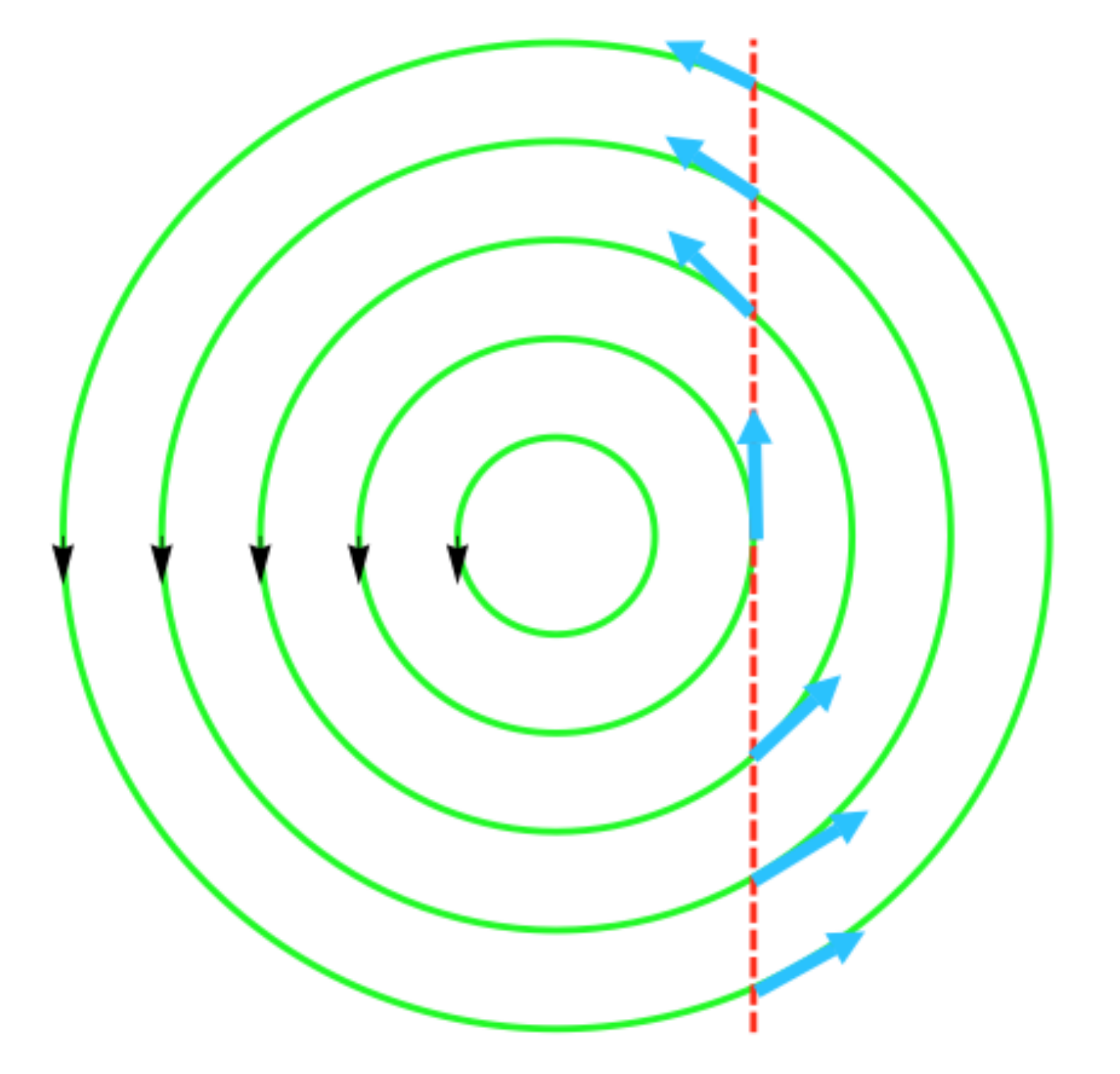}
\caption{\label{fig:fig5} (Color online) Illustration for the system described by Eq. (\ref{7.1b}). Dotted line represents
a classical trajectory of the neutral atom for a given impact parameter. Green concentric
circles represent the external magnetic field lines. Blue arrows represent the spin orientation of
the lowest energy Zeeman level for the atom. The upper Zeeman level is closed, and so
the spin of the atom is slaved to its lowest Zeeman level.} 
\end{figure}

Consider that $U$ given by Eq. (\ref{2.1}) is now
replaced by
\beq
  {U} =\exp(-i\frac{\phi}{2} { \sigma}_{3})
   \exp(i \frac{\pi}{4} { \sigma}_{1} ) \exp(i\frac{\phi}{2} { \sigma}_{3})
  \label{7.1a}
\eeq
where $ \phi $ the azimuthal angle in the $ x \, y $ plane.
Hamiltonian Eq. (\ref{1.1}) is now given
by
\beq
&& H=-\frac{\hbar^{2}}{2 m} \nabla^{2}_{\bm R} +H_{ad} \nonumber \\
&& H_{ad}={U}(\phi) H_{BO}(\rho)
 { U}^{\dag}(\phi).
\label{7.1b}
\eeq
This adiabatic Hamiltonian is identical to that of a spin 1/2 particle, with
a magnetic moment $\mu_{0}$, interacting with an external magnetic field in the $x y$ plane as shown
in Figure (\ref{fig:fig5}), i.e.
\beq
&& H_{ad} = \mu_{0} \,{\bm \sigma} \cdot {\bm B} \nonumber \\
&& {\bm B}= {\bm {\hat \phi}} \,  B(\rho).  
\label{7.1c}
\eeq
The strength of field $B(\rho)$, where $\rho$ is the distance from the origin, depends on the
physical setup. For example, a wire current passing through the origin (out of the page) generates
such a field with $ B(\rho) \sim 1/\rho$. For the purpose of this discussion we envision a setup
in which $B(\rho)$ has the constant value $ \Delta$, and then the BO energies are simply given by $\pm \Delta$, i.e.
\beq
H_{BO}= \left ( \begin{array}{cc} \Delta & 0 \\ 0 & -\Delta  \end{array} \right ) \nonumber
\eeq
and where we have set $\mu_{0}=1$.

In the adiabatic picture we again obtain Eq. (\ref{1.17}) but with the vector potential now
given by 
\beq
{ {\bm A}}  = \frac{{\bm {\hat \phi}}}{2 \rho} \left(
\begin{array}{cc}
 -1 & i \, e^{-i \phi } \\
 -i \, e^{i \phi } & 1
\end{array} \right),
\label{7.2}
\eeq
where $\phi$ is the polar angle in a cylinder coordinate system. 
The diagonal scalar potential ${ V} = H_{BO}$ contains the BO eigenvalues as its entries.
Eq. (\ref{7.2}) contains
diagonal as well as the off-diagonal components that couple the
two adiabatic channels. Suppose the channel with BO eigenvalue $+B(\rho) =\Delta$ is
closed, i.e. the collision energy $ k^{2} < \frac{2 m}{\hbar^{2}} \Delta$ as $ \rho \rightarrow \infty$.

We then ignore  
the gauge coupling to the closed channel and project onto the open
sector to obtain the approximate, one-channel, equation
\beq
&& -\frac{\hbar^{2}}{2m}
({\bm \nabla}- i {\bm A})^{2} { F}({\bm R})
+ { V}(R) { F}({\bm R})= E { F}({\bm R}), \nonumber \\
&& \label{7.3}
\eeq
where 
\beq
&& {\bm A} =\frac{{\bm {\hat \phi}}}{2 \rho} \nonumber \\
&& V(R) = -\Delta + \frac{\hbar^{2}}{8 m \rho^{2}},
\label{7.3a}
\eeq
and the second term in Eq. (\ref{7.3a}) represents the induced scalar gauge potential\cite{zyg90}.
This procedure is analogous to the one we carried out in the previous section for the 2D ``magnetic''
slab. There we showed that the BO projection is accurate in cases where the excited channel is closed,
this conclusion also applies to this system\cite{zyg10d}.
Vector potential
(\ref{7.3a}) has the property
\beq
\int_{C} d{\bm r} \cdot {\bm A} \neq 0
\label{7.5}
\eeq
where the line integral is taken along the path $C$ that encloses the origin.
Because of Eq. (\ref{7.5}) we cannot find a gauge transformation in which
the vector potential is eliminated\cite{wuyang76}. Indeed this vector potential is familiar
from Aharonov-Bohm scattering\cite{ab59}. A similar vector potential has been invoked\cite{mead76}
for poly-atomic systems in which conical intersections occur. It is note-worthy to mention that
the adiabatic eigenstates $ { U}(\theta) |\pm>$ (where $|\pm>$ are the eigenstates
of $H_{B0}$) are not multi-valued, and the BO eigenvalues $\pm B(\rho)$ are non-degenerate everywhere, so the analysis given in
Ref. \cite{mead76,mead92} does not apply here. Previous studies of geometric phase effects in reactive scattering
require the presence of conical intersections, as in studies of the hydrogen-exchange reaction\cite{marcos05}.
Unlike the example given in the previous sections, gauge potential (\ref{7.3a}) does posess singular behavior as $ \rho \rightarrow 0$,
as is evidenced by relation (\ref{7.5}) which does not vanish in the limit as ${\cal C}$ encloses and contracts upon the origin. 
Consider the following integral
\beq
W={\cal P} \exp(i \int_{{\cal C}} d{\bm R} \cdot {\bm A} )
\label{7.6} 
\eeq
where ${\cal C}$ is 
a path that starts at the point $\{\rho=R_{0},\phi=0\}$ and  follows a circular arc to the final point 
$ \{R_{0}, \phi \}$. ${\bm A}$ is given by Eq. (\ref{7.2}) and in order to evaluate this integral we
parametrize the path 
$ {\bm R}(t) = R_{0} \cos\omega t \, {\bm {\hat i}} + R_{0} \sin\omega t \, {\bm {\hat j}} $.
According to Eq. (\ref{1.4}) 
\beq
&& W(t) = T \, \exp(i \int_{0}^{t} dt' A(t') ) \nonumber \\ 
&& A(t') = \frac{d {\bm R}}{dt'} \cdot {\bm A} = \frac{\omega}{2} \left ( \begin{array}{cc} -1 & i \exp(-i \omega t) \\
-i \exp(i \omega t) & 1 \end{array} \right ) \nonumber \\
\label{7.7}
\eeq
where $\omega t = \phi$.
Now,
\beq
\frac{d W}{dt} = i A(t) W
\eeq
which integrated, gives
\beq
W(t) = \frac{1}{\sqrt{2}} \left ( \begin{array}{cc} 1 & -i \exp(-i \omega t) \\ -i \exp(i \omega t) & 1 \end{array} \right ).
\label{7.8}
\eeq
Indeed we find, replacing $t$ with $\phi/\omega$, that $W(t) = U^{\dag}$, given in Eq. (\ref{7.1a}). 
When path ${\cal C}$ makes a closed circuit we find $ W(2\pi/\omega)=1$ as it must. 
Therefore ${\bm A}$ is a pure gauge and is not singular.  However its diagonal, Abelian, components $\pm \frac{{\bm {\hat \phi}}}{2 \rho} $ need not be and, 
in this case, lead to the singular behavior evidenced by relation (\ref{7.5}).
At low collision energies, quantum dynamics effectively projects out the channel in which this manifestation of a
non-integrable phase factor emerges.

Though vector potential Eq. (\ref{7.3a}) is nontrivial it does not lead to an effective Lorentz force for $\rho \neq 0$,
nevertheless its presence has a profound effect on the scattering properties of this system\cite{zyg10d}.
These examples illustrate how gauge potentials arise due to the ``dressing'' of the atom (a spin 1/2 system) by 
 external fields. Below we show how this relates to a central question of geometric magnetism,
how and why do non-trivial gauge potentials arise in inter-atomic interactions?
 We now take our system to be that of two spin - 1/2 systems that
 interact via a spin-spin
dipolar interaction. 
Such systems are often realized in nature, including that for the interaction of two hydrogen atoms \cite{zyg09,zyg10d}.
  They also serve as a template for a non-trivial two-qubit quantum gate\cite{zyg09}. 
Here the adiabatic energy
splittings are generated by the internal spin-spin interaction, and the BO energy matrix is sandwiched by
rotation operators similar to that given above. In an adiabatic expansion (now a four channel, or two-qubit problem) the
PSS equations possess a four channel vector potential, similar to those given above and discussed in detail below. 
\section{Internal gauge potentials}
Consider Hamiltonian (\ref{7.1c}) in three spatial dimensions and
with ${\bm B}$ spherically symmetric about the origin. This (or closely related to it) system 
has been discussed extensively in the literature\cite{ber84,stone86,ber93}.
In Berry's model \cite{ber84}, ${\bm R}$ is treated as a classical variable, Stone quantized
${\bm R}$ and considered the full quantal solutions.
Berry and Robbins\cite{ber93} also
studied, from a semi-classical perspective, the role of the effective Lorentz forces\cite{zyg87a,ber89} that arise
in such systems.
The diagonal 
components of ${\bm A}$ produce the effective magnetic field of a Dirac magnetic monopole\cite{dirac31,moo86}. These
gauge potentials are realized in molecular and atomic systems\cite{moo86,zyg92} and
Stone \cite{stone86} suggested a ``Heath-Robinson-type device'', a rotating solenoid
interacting with a spin 1/2 particle, to serve as a physical realization.

Consider the Hamiltonian for two, interacting, spin-1/2 particles\cite{zyg09} with magnetic moments
\beq
H=-\frac{\hbar^{2}}{2m}
\nabla^{2}_{\bm R}+H_{ad}({\bm R})
\label{8.1}
\eeq
where ${\bm R} \equiv (R,\theta,\phi) $ is the 3D inter-atom separation vector expressed in polar coordinates.
The adiabatic Hamiltonian is given by \cite{zyg10b}
\beq
&& H_{ad} = H_{0}+H_{dip} \nonumber \\
&& H_{0} = \Bigl (   {^{3}\Sigma(R)}- ^{1}\Sigma(R) \Bigr )
         {\bm S}_{a} \cdot {\bm S}_{b}  + \frac{3 {^{3}\Sigma(R)}+^{1}\Sigma(R)}{4} 
	 	 \nonumber \\
&& H_{dip} =\frac{\alpha^{2} }{R^{3}} \Bigl [
        {\bm S}_{a} \cdot  {\bm S}_{b}  - 3  
        \frac{({\bm S}_{a} \cdot  {\bm R}) ({\bm S}_{b} \cdot {\bm R})}{R^{2}} \Bigr ].	 
\label{8.2}
\eeq  
The scalar potentials, ${^{3}\Sigma(R)}, \, ^{1}\Sigma(R)$, are arbitrary and depend on the polarization
properties of the constituents. For a structure-less system, e.g. two electrons, these scalar terms
are degenerate and $ H_{0}$ reduces to the long-range Coulomb interaction. In the case of two neutral
spin 1/2 atoms, such as two Alkali atoms, $H_{0}$ (in atomic units) represents the isotropic electrostatic interaction
between the atoms and ${^{3}\Sigma(R)}$ and $^{1}\Sigma(R)$ are the triplet and singlet BO energies of the electrostatic
Hamiltonian respectively. $ S_{i}$ are spin-1/2 operators.
$H_{dip}$ is the long range component of the anisotropic
magnetic interaction, $\alpha$ is a constant and for two electrons is
given by the fine structure constant when atomic units are employed. We re-express \cite{zyg09,zyg10} 
\beq
H_{ad} = U(\theta \phi) H_{BO}(R)  U^{\dag}(\theta \phi)
\label{8.3}
\eeq
where
\beq
&& U= U_{a} \otimes U_{b} \, Z \nonumber \\
&& U_{a}=U_{b} =\exp(-i { \sigma}_{3} \phi) \exp(-i { \sigma}_{2} \theta)
 \exp(i { \sigma}_{3} \phi) \nonumber \\
&& Z= \left ( \begin{array}{cccc} 0 & 0 & 1 & 0 \\
                               0 & - \frac{1}{\sqrt{2}} &0  & - \frac{1}{\sqrt{2}}  \\
			       0 & -\frac{1}{\sqrt{2}}  &0  & \frac{1}{\sqrt{2}}  \\
			       1 & 0 & 0 & 0 
			        \end{array} \right )
\label{8.3a}
\eeq
and 
\begin{widetext}
\beq
H_{BO}= \left ( \begin{array}{cccc}  {^{3}\Sigma(R)}-\frac{\alpha^{2}}{2 R^{3}} & 0 & 0 & 0 \\
                               0 &  {^{3}\Sigma(R)}+\frac{\alpha^{2}}{ R^{3}} & 0 & 0 \\
			       0 & 0 & {^{3}\Sigma(R)}-\frac{\alpha^{2}}{2 R^{3}} & 0 \\
			       0 & 0 & 0 & {^{1}\Sigma(R)} \end{array} \right )
\eeq
\end{widetext}
is a function of the internuclear distance $R$ only.
Using definition (\ref{1.8}) we find that the induced vector potential
has the block form,
\begin{widetext}
\beq
&& {\bm {A}}= \left ( \begin{array}{cc} {\bm { A}_{t}} & 0 \\
             0 & 0 \end{array} \right )   \label{4.3b} \\
&& {\bm { A}_{t}} = \frac{1}{R}
 \left (
 \begin{array}{ccc}
 \hat{ {\bm \phi}} \, \tan[\frac{\theta}{2}]   & 
  \frac{\exp(i \phi)}{\sqrt{2}}( -i \hat{\bm \theta} + \hat{\bm \phi})  & 0 \\
  \frac{\exp(-i \phi)}{\sqrt{2}}( i \hat{\bm \theta} + \hat{\bm \phi})  & 0 & 
  \frac{\exp(i \phi)}{\sqrt{2}}( -i \hat{\bm \theta} + \hat{\bm \phi})  \\
 0 & \frac{\exp(-i \phi)}{\sqrt{2}}( i \hat{\bm \theta} + \hat{\bm \phi})  & 
 -\hat{ {\bm \phi}} \, \tan[\frac{\theta}{2}] 
 \end{array}
 \right ).
 \nonumber
 \eeq
 \end{widetext}
 We note the following features
  \cite{zyg09,zyg10}
 \begin{itemize}
 \item The degenerate triplet sector of the diagonal scalar potential is split due to the inclusion
 of $H_{dip}$. The states
 corresponding to magnetic quantum numbers $|M|=1$ contain an attractive component 
 $ -\alpha^{2}/2R^{3}$, whereas the $M=0$ state contains a repulsive component $ \alpha^{2}/2R^{3} $.
 \item Non-trivial vector gauge potentials are induced. They contain off-diagonal components that
 couple the $ M=\pm 1$ adiabatic channels with the $ M=0 $ channel. There is no direct coupling between
 the two $ M=\pm 1$ channels, and the diagonal components are that of  Dirac-magnetic monopoles
 of integer charges $\pm 1$.
 \item Because the energies of the diagonal entries in $H_{BO}$ are split, a gauge transformation cannot be found
in which the temporal components of the field strength tensor vanish\cite{zyg87a}. As a consequence the gauge potentials are non-trivial
in the sense described for the systems discussed in the previous sections.
\end{itemize}
 When solving the scattering equations in the adiabatic gauge the induced gauge potentials, depending on the collision energy,  contribute 
 to spin changing cross sections. Though exact scattering calculations for such a system (i.e. two ground state hydrogen atoms) have
 been performed in the diabatic gauge\cite{zyg10} the question of how and in what manner the induced gauge potentials influence
 scattering cross sections remains to be investigated\cite{zyg09}.  
\section{Summary and Discussion}
In molecular physics and atomic collision theory one is typically faced with
a many-body Hamiltonian having the structure given by Eq. (\ref{1.1}). The adiabatic Hamiltonian
$ H_{ad}({\bm R},{\bm r}) $ is a function of the ``fast degrees'' of freedom ${\bf r}$ as well as
${\bm R}$, the coordinates for the positions of the atoms. Using the eigenstates of $H_{ad}$ as a basis
one arrives at the set of Equations (\ref{0.3}) in which the matrix vector gauge potential ${\bm A}$ manifests.
In applications, Eq. (\ref{0.3}) involves an infinite set of coupled equations and a truncation procedure
must be invoked in order to proceed with a working theory. 
The remaining gauge potentials, after truncation, are non-trivial in the
sense that the resulting spatial field strength tensor does not vanish\cite{zyg87a} and, as a consequence, lead to effective
gauge forces. They emerge in addition to the traditional Hellmann-Feynman force. We introduced and studied
the gauge structure of several low-dimensional scattering systems that are solvable and where it was not necessary to employ
the truncation, or BO, approximation. For the adiabatic Hamiltonian we replaced the fast degrees of freedom
with the discrete Pauli coordinates ${\bm \sigma}$,
\beq
H_{ad}({\bm R},{\bm r}) \rightarrow H_{ad}({\bm R},{\bm \sigma}) 
\nonumber
\eeq
so that a complete set of basis functions can be
used to obtain the exact solutions of the PSS equations (\ref{0.3}). Though our choices for the adiabatic
 Hamiltonians varied in complexity they shared the feature that
\beq
H_{ad} = U({\bm R}) H_{BO} U^{\dag}({\bm R})
\eeq
where $U$ is a unitary operator which can be characterized by it's gauge field as given in Eq. (\ref{1.3}).
$H_{BO}$ is a diagonal matrix that contains the energy eigenvalues for $H_{ad}$, and in our discussion we
were careful to choose systems in which $H_{BO}$ did not feature degeneracies at physical values for the slow
parameter ${\bm R}$. The energy gap in $H_{BO}$ explictly ``breaks'' the symmetry inherent in the definition
of $U$. The matrix structure of $H_{ad}$ depends on the particular system. For example, in our
discussion relating to Figure (\ref{fig:fig1}) $H_{ad}$ can be  ``engineered''\cite{dal10b} by the application of a laser field on a two-level system,
for the system described in Figure (\ref{fig:fig5}) $H_{ad}$ is generated by an external magnetic field coupled to a neutral spin-1/2 particle.
In Eq. (\ref{8.3a}) $U$ is a rotation operator that characterizes the dipolar interaction Eq. (\ref{8.2}) for two spin-1/2 particles.

For the system described by Eq. (\ref{2.2}) we were able to solve for the scattering parameters in both the adiabatic and diabatic
descriptions and showed that the quantity,
\beq
{\cal F} \equiv \int_{\cal C} \, d {\bm s} \cdot {\bm j}  
\eeq
where ${\bm j}$ is a current, is invariant with respect to the choice in representation. Furthermore, for the closed loop
${\cal C}$ shown in Figure (\ref{fig:fig4}) we found that ${\cal F}$ does not vanish in the low-energy limit. In the
adiabatic representation the non-vanishing contribution arises from the gauge potential ${\bm A}$ that define $U$.
 In the low-energy limit, where the excited adiabatic channel is closed, we found that the BO projection procedure
serves as a good approximation for solutions obtained by the fully coupled calculation. The effective single-channel
gauge potential predicts a ``magnetic'' field, given by Eq. (\ref{3.5}), that leads to the appearance of an effective
Lorentz force on the scattered atom. The BO amplitude which is coupled to the, non-trivial, gauge potential allows the BO
current to satisfy the condition ${\cal F} \neq 0$. We compared the quantal scattering solutions of this system  with
that of a charged particle interacting with a ferromagnetic slab, and we found that the former predicts, according to Eq. (\ref{3.9a}), a deflection
angle in harmony with that predicted by the latter. Though the multichannel gauge potential ${\bm A}$ is trivial in the sense that the
spatial components of the curvature vanish, in accordance with Eq. (\ref{1.5}), non-trivial gauge behavior manifests in the low energy limit.
The adiabatic Hamiltonian does not suffer from any degeneracies, and we propose
that the non-trivial gauge behavior is akin to the phenomenon of symmetry breaking\cite{nambu61}. 
This system serves as a valuable template for more complex examples, but it also highlights possible
applications for geometric magnetism. For example, in section III.E we showed how these ideas can be applied toward
the engineering of an effective ``magnetic'' lens for neutral atoms.       

We showed the relationship between systems in which dressing is carried out by external agents, 
such as a magnetic field for the system
described by Eq. (\ref{7.1b}), or a laser field\cite{dal10b} for the two level system given by Eq. (\ref{2.1}),
with those in which it is accomplished by mutual inter-particle interactions.
For the interacting two-qubit system give by Eqs. (\ref{8.3})
and (\ref{8.3a})  ``dressing''  is accomplished by the 2-body tensor interaction Eq. (\ref{8.2}). 
We argue that, regardless of the nature of the dressing agent, gauge behavior manifest in these disparate
systems via the same mechanism, and we do not advocate the use different labels,
 e.g. fictitious, artificial, synthetic or geometric magnetism,  to identify the phenomena. 
 We propose that geometric magnetism is an appropriate terminology that underscores its universality. 

\appendix
\section{Coupled channel solution}
We compare the solution of the approximate adiabatic solution Eq. (\ref{1.20}) to that obtained by
solving the coupled equations (\ref{1.17}). For the sake of economy in notation we set $\hbar=1$ in this appendix. 
In region $ x>0$ we take the ansatz 
\beq
{ {\underline F}}_{ad} = \left(
\begin{array}{c}
 F_{c} \\
 F_{o}
\end{array} 
\right) = \exp(i \Omega \, x) \left(
\begin{array}{c}
 c_{c} \\
 c_{o}
\end{array}
\right).
\label{1.21a}
\eeq
Inserting this ansatz into Eq. (\ref{1.17}) we obtain a set of homogeneous linear equations for the
coefficients $c_{o}, c_{c}$ whose non-trivial solutions require that $\Omega$ obey the eigenvalue
equations
\beq
&& -\Omega^{2}-A_{0}^{2}-A_{1}^{2} + 2 m E \mp  \nonumber \\
&& \sqrt{(2 m \Delta-2 A_{0} \Omega)^{2}+ (2 A_{1}\Omega)^2} =0. \nonumber \label{1.22a}
\eeq
Analytic expressions for the eigenvalues can be gleaned using the formula of Ferraria\cite{carpenter} for the roots of the quartic
equation. An outline for calculating the solutions is given in the Supplement.
 A simpler and transparent expression that is valid in the, adiabatic, limit where $m \rightarrow \infty$  is sufficient for our
purposes. In this limit the roots are given by 
\beq
&& \Omega_{c} \approx A_{0} \pm 2 i \sqrt{m \Delta} \nonumber \\
&& \Omega_{o}  \approx -A_{0} \pm \sqrt{k^{2}-A_{1}^{2}} .
\label{1.22b}
\eeq
where we set $ E = - \Delta + \frac{\hbar^{2} k^{2}}{2 m} $ and  require that $  \frac{\hbar^{2} k^{2}}{2 m} < 2\Delta $ i.e.
the collision energy is much smaller than the energy defect between the two BO potentials. We then
obtain the following, approximate, solutions to Eq. (\ref{1.17}), for $x>0$,
\beq
&& {\underline F}_{c} = \exp(i A_{0} x) \exp(\pm\sqrt{4 m \Delta} \, x) \times 
\left ( \begin{array}{c}  \pm i \\ \frac{A_{1}}{\sqrt{m \Delta}} \end{array} \right )
\label{1.23a}
\eeq
and
\beq
{\underline F}_{o} = \exp(-i A_{0} x) \exp(\pm i \, \sqrt{k^{2}-A_{1}^{2}}  \, x) 
\left ( \begin{array}{c} 0 \\ 1  \end{array} \right ).
\label{1.24a}
\eeq
In the region $ x <0 $ we obtain
\beq
{\underline F}   = \left ( \begin{array}{c}  S_{co} \, \exp(\sqrt{4 m \Delta} \, x) \\ \exp(i k (x-L)) + R \exp(-i k (x-L)) \end{array} \right )
\eeq
and solve for the open channel reflection coefficient $R$ by requiring the solution at $ x>0$ to match the boundary condition at $x=L$, and
then matching it with the asymptotic solutions at $x=0$. However, care must be taken when matching the derivatives of the adiabatic amplitudes
since the vector potential ${ A}$ is discontinuous across the boundary.
We require the diabatic amplitudes given in Eq. (\ref{1.14a}) and their derivatives to be continuous across the boundary. However since the adiabatic
amplitudes are related to them according to Eq. (\ref{1.16}) we require that
\beq
{{\underline F}'_{ad}}_{>}-{{\underline F}'_{ad}}_{<} = i \, { A} \, {\underline F}_{ad}(0)
\label{1.25a}
\eeq
where $ {{\underline F}'_{ad}}_{>}$, ${{\underline F}'_{ad}}_{<}$ are the derivatives of the amplitudes in the regions $x=0\pm \epsilon$ respectively,
and  ${\underline F}_{ad}(0)$ is the value of the amplitude at $x=0$. Matching the amplitudes and enforcing condition
Eq. (\ref{1.25a}) we obtain, as $k \rightarrow 0$,
\beq
R = -1 + 2 i k \, \Bigl (L -\frac{\tanh(A_{1} L)}{A_{1}} \Bigr ) \, + \, {\cal O}(k^{2}).
\label{1.26a}
\eeq
In the limit where $\Delta=0$ the exact eigenvalues $\Omega = \pm k \pm \sqrt{A_{0}^{2}+A_{1}^{2}}$, allow
simple analytic expressions for solutions of Eq. (\ref{1.17}). Though the eigensolutions are non-trivial (see Supplementary material) in the region $ x>0$,
carrying out the procedure that resulted in Eq. (\ref{1.26a}) leads to $ R=-1$ as it must. 

\section{Scattering by a 2D ferromagnetic strip}
\begin{figure}[ht]
\centering
\includegraphics[width=0.9\linewidth]{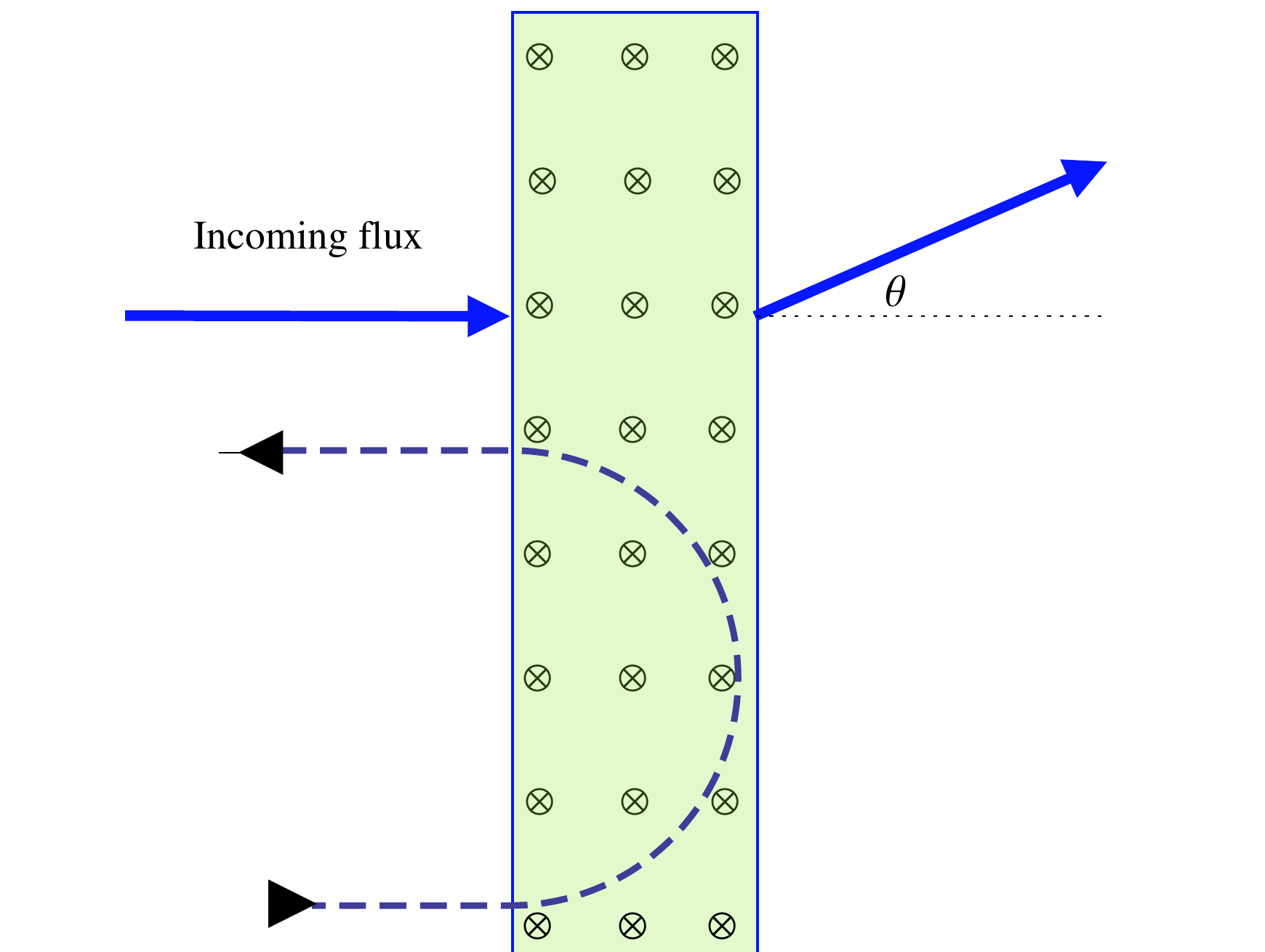}
\caption{\label{fig:fig1A} (Color online) Normal incidence of a charged particle on a ferromagnetic slab of width $L$.
 The magnetic field ${\bm B}$ is directed into the
 page and the dashed line depicts a typical classical trajectory. 
 In this figure the classical trajectory describes
 a particle that does not have the required velocity to overcome the barrier shown in Fig. \ref{fig:fig1} }
\end{figure}

We consider the scattering of a charged particle (here we set $\frac{q}{\hbar c} =1$) by a 2D ferromagnetic slab as 
shown in Figure \ref{fig:fig1A}.
The particle with energy $E$ is traveling to the right, along the $x$-direction, and impacts the slab, in the normal direction, 
of thickness $L$. We consider the Schrodinger equation
\beq
-\frac{\hbar^{2}}{2m} \Bigl ( {\bm \nabla} - i {\bm A} \Bigr )^{2} \psi = E \psi.
\label{a.0}
\eeq
The vector potential ${\bm A}= A \, {\bm {\hat j}} $ is given by
\beq
&& A=  B_{0} \theta(x+L/2)(x+L/2)\theta(L/2-x)+ \nonumber \\
&&  \theta(x-L/2) L B_{0}  \label{a.1} 
\eeq
and  
\beq
&&{\bm B}= {\hat {\bm z}} \Bigl (
\frac{\partial A_{y}}{\partial x} -  \frac{\partial A_{x}}{\partial y} \Bigr ) = \nonumber \\
&& {\hat {\bm z}} \, B_{0} \theta(x+L/2)\theta(L/2-x). 
\eeq
We assume the particle approaches from the left (for $x < -L/2$ ) with the incident
wave boundary condition,
\beq
\psi = \exp(i k x) + r \, \exp(-i k x) 
\label{a.1a}
\eeq  
where $ k = \sqrt{\frac{2 m}{\hbar^{2}} E}$ and $r$ is the reflection coefficient. 
In the region $ -L/2 < x < L/2 $ the Schrodinger equation becomes,
 \beq
 \frac{\partial^{2} \psi}{\partial x^{2}} - B_{0}^{2} (x+L/2)^{2} \psi+k^{2} \psi=0
 \label{a.2}
 \eeq
 whose solutions are,
  \beq
 && c_{1} D_{\nu_{1}}\left(\sqrt{B_{0}/2} (L+2 x) \,\right) +  c_{2}  D_{\nu_{2} }\left(i \sqrt{B_{0}/2} (L+2 x) \, \right).
   \nonumber \\ \label{a.3}
 \eeq
 where
 \beq
 && \nu_{1} \equiv \frac{k^{2}}{2 B_{0}} -\frac{1}{2} \nonumber \\
 && \nu_{2} \equiv  -\frac{k^{2}}{2 B_{0}} -\frac{1}{2}\nonumber
 \eeq
 and $ D_{\nu}(x)$ are the Parabolic Cylinder Functions\cite{abrom}.
 For $ x>L/2$ the amplitudes satisfy,
 \beq
 &&\frac{\partial^{2} \psi}{\partial x^{2}} - \Phi^{2} \psi+k^{2} \psi=0 \nonumber \\
 && \Phi = B_{0} L
 \label{a.4}
 \eeq
 and whose form is determined by the scattering boundary conditions
 \beq
 \psi(x) = t \, \exp(i \sqrt{k^{2} - \Phi^{2}} x)   \quad k> |\Phi| \nonumber \\
 \psi(x) = t \, \exp(-\sqrt{|k^{2} - \Phi^{2}|} x)   \quad k< |\Phi| \nonumber \\
 \label{a.5}
 \eeq
\subsection{Reflection and transmission coefficients}
We now solve for reflection and transmission coefficients.
We take the logarithmic derivative, at $x=L/2$, of the internal solution Eq. (\ref{a.3}) and match  
it with that obtained using the asymptotic boundary condition Eq. (\ref{a.5}) resulting
in the ratio,
\beq
&& \gamma \equiv c_{2}/c_{1}= \nonumber \\
&& \frac{D'_{\nu_{1}}\left(\sqrt{2 B_{0}}\,L \right)-
 i \, \sqrt{k^{2}-\Phi^{2} } \, D_{\nu_{1}}\left(\sqrt{2 B_{0}}\,L\right)}
 {-D'_{\nu_{2}}\left(i \sqrt{2 B_{0}}\, L\right) + 
 i \, \sqrt{k^{2}-\Phi^{2}} \, D_{\nu_{2}}\left(i \sqrt{2 B_{0}}\, L\right)}. 
\label{a.6}
\eeq
Taking the incident wave and matching it, and its derivative, with the internal solution we obtain
\beq
&& r = \exp(-i L k) \frac{k+i \, y}{k-i \, y} \nonumber \\
&& y= \frac{D'_{\nu_{1}}+ \gamma \, D'_{\nu_{2}}}
{D_{\nu_{1}}+ \gamma \, D_{\nu_{2}} }
\label{a.8}
\eeq

and
\beq
&& t  = 2 i k \exp(-i \frac{L k}{2}) \exp(-\frac{i L}{2} \,\sqrt{k^{2}-\Phi^{2}})
\times \nonumber \\
&& \frac{{ \Bigl ( D_{\nu_{1}}\left(\sqrt{2 B_{0}} \, L \right) + \gamma \,
   D_{\nu_{2}}\left(i \sqrt{2 B_{0}} \, L \right) \Bigr ) }}
{D'_{\nu_{1}}+
\gamma  \, D'_{\nu_{2}} + 
i k \Bigl ( D_{\nu_{1}}+
\gamma \, D_{\nu_{2}} \Bigr )} 
\label{a.9}
\eeq
where we used the shorthand $ D_{\nu} \equiv D_{\nu}(0)$.

Consider the limit $ L \rightarrow 0 $ as the flux density $ \Phi \equiv B_{0} L $ stays constant.
We find, in this limit,
\beq
\gamma \rightarrow i + {\cal O} (\sqrt{L})
\eeq
as
\beq
y \rightarrow i \sqrt{k^{2}-\Phi^{2}}.
\eeq
Thus
\beq
&& r \rightarrow \frac{k-\sqrt{k^{2}-\Phi^{2}}}{k+\sqrt{k^{2}-\Phi^{2}}}  \nonumber \\
&& t = 1 + r
\eeq
and are the scattering parameters obtained for a system in which a particle is incident on 
a step function potential of height $ \Phi^{2} $ at $ x>0$.
\subsection{Conserved currents}
For the Schrodinger equation
\beq
i \hbar \frac{\partial \psi}{\partial t} = -\frac{\hbar^{2}}{2m} \Bigl (
{\bm \nabla}- i {\bm A} \Bigr )^{2} \psi + \frac{\hbar^{2} k^{2}}{2m} \psi 
\label{a.10}
\eeq
the current density is given by
\beq
{\bm j}= \frac{i \hbar}{2m} ( \psi {\bm \nabla} \psi^{*} - \psi^{*} {\bm \nabla} \psi +2 i {\bm A} \psi^{*} \psi ).
\label{a.11}
\eeq
Inserting expression Eq. (\ref{a.1.a}) into Eq. (\ref{a.11}) we obtain
\beq
j_{x} = \frac{\hbar k}{m} \Bigl ( 1 -|r|^{2} \Bigr )
\label{a.12}
\eeq
which can be expressed as incident $  j_{in} = \frac{\hbar k}{m}$ and reflected 
$j_{r} = -\frac{\hbar k}{m} |r|^{2} $ currents, respectively. In the region $ x> L/2$ we find
\beq
&& j_{x} =\frac{\hbar}{m} |t|^{2} \sqrt{k^{2}-\Phi^{2}} \nonumber \\
&& j_{y} =  - \frac{\hbar}{m} |t|^{2} \Phi.  
\eeq 
where we have used Eqs. (\ref{a.1}) and (\ref{a.5}).
Thus the transmitted current makes an angle $\theta$ with respect to the normal given
\beq
\tan\theta \equiv \frac{j_{y}}{j_{x}}= -\frac{ \Phi }{\sqrt{k^{2}-\Phi^{2}}}
\eeq
where $ \Phi = B_{0} L$, and agrees with the classical angle of deflection.

% ACKNOWLEDGMENTS COME HERE: 
\acknowledgments
 This work was supported, in part, by NSF grant PHY-0758140. I wish to thank an anonymous reviewer for helpful suggestions.

%\bibliographystyle{apsrev4-1}
%\bibliography{newbib}

\begin{thebibliography}{48}%
\makeatletter
\providecommand \@ifxundefined [1]{%
 \@ifx{#1\undefined}
}%
\providecommand \@ifnum [1]{%
 \ifnum #1\expandafter \@firstoftwo
 \else \expandafter \@secondoftwo
 \fi
}%
\providecommand \@ifx [1]{%
 \ifx #1\expandafter \@firstoftwo
 \else \expandafter \@secondoftwo
 \fi
}%
\providecommand \natexlab [1]{#1}%
\providecommand \enquote  [1]{``#1''}%
\providecommand \bibnamefont  [1]{#1}%
\providecommand \bibfnamefont [1]{#1}%
\providecommand \citenamefont [1]{#1}%
\providecommand \href@noop [0]{\@secondoftwo}%
\providecommand \href [0]{\begingroup \@sanitize@url \@href}%
\providecommand \@href[1]{\@@startlink{#1}\@@href}%
\providecommand \@@href[1]{\endgroup#1\@@endlink}%
\providecommand \@sanitize@url [0]{\catcode `\\12\catcode `\$12\catcode
  `\&12\catcode `\#12\catcode `\^12\catcode `\_12\catcode `\%12\relax}%
\providecommand \@@startlink[1]{}%
\providecommand \@@endlink[0]{}%
\providecommand \url  [0]{\begingroup\@sanitize@url \@url }%
\providecommand \@url [1]{\endgroup\@href {#1}{\urlprefix }}%
\providecommand \urlprefix  [0]{URL }%
\providecommand \Eprint [0]{\href }%
\providecommand \doibase [0]{http://dx.doi.org/}%
\providecommand \selectlanguage [0]{\@gobble}%
\providecommand \bibinfo  [0]{\@secondoftwo}%
\providecommand \bibfield  [0]{\@secondoftwo}%
\providecommand \translation [1]{[#1]}%
\providecommand \BibitemOpen [0]{}%
\providecommand \bibitemStop [0]{}%
\providecommand \bibitemNoStop [0]{.\EOS\space}%
\providecommand \EOS [0]{\spacefactor3000\relax}%
\providecommand \BibitemShut  [1]{\csname bibitem#1\endcsname}%
\let\auto@bib@innerbib\@empty
%</preamble>
\bibitem [{\citenamefont {Feynman}(1939)}]{feyn39}%
  \BibitemOpen
  \bibfield  {author} {\bibinfo {author} {\bibfnamefont {R.~P.}\ \bibnamefont
  {Feynman}},\ }\href@noop {} {\bibfield  {journal} {\bibinfo  {journal} {Phys.
  Rev.}\ }\textbf {\bibinfo {volume} {56}},\ \bibinfo {pages} {340} (\bibinfo
  {year} {1939})}\BibitemShut {NoStop}%
\bibitem [{\citenamefont {{Dalgarno}}\ and\ \citenamefont
  {{McCarroll}}(1956)}]{dal56}%
  \BibitemOpen
  \bibfield  {author} {\bibinfo {author} {\bibfnamefont {A.}~\bibnamefont
  {{Dalgarno}}}\ and\ \bibinfo {author} {\bibfnamefont {R.}~\bibnamefont
  {{McCarroll}}},\ }\href {\doibase 10.1098/rspa.1956.0184} {\bibfield
  {journal} {\bibinfo  {journal} {Royal Society of London Proceedings Series
  A}\ }\textbf {\bibinfo {volume} {237}},\ \bibinfo {pages} {383} (\bibinfo
  {year} {1956})}\BibitemShut {NoStop}%
\bibitem [{\citenamefont {Moody}\ \emph {et~al.}(1986)\citenamefont {Moody},
  \citenamefont {Shapere},\ and\ \citenamefont {Wilczek}}]{moo86}%
  \BibitemOpen
  \bibfield  {author} {\bibinfo {author} {\bibfnamefont {J.}~\bibnamefont
  {Moody}}, \bibinfo {author} {\bibfnamefont {A.}~\bibnamefont {Shapere}}, \
  and\ \bibinfo {author} {\bibfnamefont {F.}~\bibnamefont {Wilczek}},\
  }\href@noop {} {\bibfield  {journal} {\bibinfo  {journal} {Phys. Rev. Lett.}\
  }\textbf {\bibinfo {volume} {56}},\ \bibinfo {pages} {893} (\bibinfo {year}
  {1986})}\BibitemShut {NoStop}%
\bibitem [{\citenamefont {Shapere}\ and\ \citenamefont
  {Wilczek}(1989)}]{shap89}%
  \BibitemOpen
  \bibfield  {author} {\bibinfo {author} {\bibfnamefont {A.}~\bibnamefont
  {Shapere}}\ and\ \bibinfo {author} {\bibfnamefont {F.}~\bibnamefont
  {Wilczek}},\ }\href@noop {} {\emph {\bibinfo {title} {Geometric Phases in
  Physics}}}\ (\bibinfo  {publisher} {World Scientific Publishing Company},\
  \bibinfo {year} {1989})\BibitemShut {NoStop}%
\bibitem [{\citenamefont {Zygelman}(1987)}]{zyg87a}%
  \BibitemOpen
  \bibfield  {author} {\bibinfo {author} {\bibfnamefont {B.}~\bibnamefont
  {Zygelman}},\ }\href@noop {} {\bibfield  {journal} {\bibinfo  {journal}
  {Phys. Lett. A}\ }\textbf {\bibinfo {volume} {125}},\ \bibinfo {pages} {476}
  (\bibinfo {year} {1987})}\BibitemShut {NoStop}%
\bibitem [{\citenamefont {Berry}(1989)}]{ber89}%
  \BibitemOpen
  \bibfield  {author} {\bibinfo {author} {\bibfnamefont {M.~V.}\ \bibnamefont
  {Berry}},\ }in\ \href@noop {} {\emph {\bibinfo {booktitle} {Geometric Phases
  in Physics}}},\ \bibinfo {editor} {edited by\ \bibinfo {editor}
  {\bibfnamefont {A.}~\bibnamefont {Shapere}}\ and\ \bibinfo {editor}
  {\bibfnamefont {F.}~\bibnamefont {Wilczek}}}\ (\bibinfo  {publisher} {World
  Scientific Publishing Company},\ \bibinfo {year} {1989})\ p.~\bibinfo {pages}
  {1}\BibitemShut {NoStop}%
\bibitem [{\citenamefont {{Zygelman}}(1990)}]{zyg90}%
  \BibitemOpen
  \bibfield  {author} {\bibinfo {author} {\bibfnamefont {B.}~\bibnamefont
  {{Zygelman}}},\ }\href {\doibase 10.1103/PhysRevLett.64.256} {\bibfield
  {journal} {\bibinfo  {journal} {Physical Review Letters}\ }\textbf {\bibinfo
  {volume} {64}},\ \bibinfo {pages} {256} (\bibinfo {year} {1990}). \bibinfo {note} {We correct a typo in Eq. (7) of
  that paper. A factor of $R$ should be inserted in the denominator of that expression}  }\BibitemShut
  {NoStop}%
\bibitem [{\citenamefont {Bliokh}\ and\ \citenamefont {Bliokh}(2005)}]{bli05}%
  \BibitemOpen
  \bibfield  {author} {\bibinfo {author} {\bibfnamefont {K.~Y.}\ \bibnamefont
  {Bliokh}}\ and\ \bibinfo {author} {\bibfnamefont {Y.~P.}\ \bibnamefont
  {Bliokh}},\ }\href@noop {} {\bibfield  {journal} {\bibinfo  {journal} {Annals
  of Physics}\ }\textbf {\bibinfo {volume} {319}},\ \bibinfo {pages} {13}
  (\bibinfo {year} {2005})}\BibitemShut {NoStop}%
\bibitem [{\citenamefont {{Lin}}\ \emph
  {et~al.}(2009{\natexlab{a}})\citenamefont {{Lin}}, \citenamefont {{Compton}},
  \citenamefont {{Perry}}, \citenamefont {{Phillips}}, \citenamefont
  {{Porto}},\ and\ \citenamefont {{Spielman}}}]{lin09a}%
  \BibitemOpen
  \bibfield  {author} {\bibinfo {author} {\bibfnamefont {Y.}~\bibnamefont
  {{Lin}}}, \bibinfo {author} {\bibfnamefont {R.~L.}\ \bibnamefont
  {{Compton}}}, \bibinfo {author} {\bibfnamefont {A.~R.}\ \bibnamefont
  {{Perry}}}, \bibinfo {author} {\bibfnamefont {W.~D.}\ \bibnamefont
  {{Phillips}}}, \bibinfo {author} {\bibfnamefont {J.~V.}\ \bibnamefont
  {{Porto}}}, \ and\ \bibinfo {author} {\bibfnamefont {I.~B.}\ \bibnamefont
  {{Spielman}}},\ }\href {\doibase 10.1103/PhysRevLett.102.130401} {\bibfield
  {journal} {\bibinfo  {journal} {Physical Review Letters}\ }\textbf {\bibinfo
  {volume} {102}},\ \bibinfo {pages} {130401} (\bibinfo {year}
  {2009}{\natexlab{a}})}\BibitemShut {NoStop}%
\bibitem [{\citenamefont {{Lin}}\ \emph
  {et~al.}(2009{\natexlab{b}})\citenamefont {{Lin}}, \citenamefont {{Compton}},
  \citenamefont {{Jim{\'e}nez-Garc{\'{\i}}a}}, \citenamefont {{Porto}},\ and\
  \citenamefont {{Spielman}}}]{lin09b}%
  \BibitemOpen
  \bibfield  {author} {\bibinfo {author} {\bibfnamefont {Y.}~\bibnamefont
  {{Lin}}}, \bibinfo {author} {\bibfnamefont {R.~L.}\ \bibnamefont
  {{Compton}}}, \bibinfo {author} {\bibfnamefont {K.}~\bibnamefont
  {{Jim{\'e}nez-Garc{\'{\i}}a}}}, \bibinfo {author} {\bibfnamefont {J.~V.}\
  \bibnamefont {{Porto}}}, \ and\ \bibinfo {author} {\bibfnamefont {I.~B.}\
  \bibnamefont {{Spielman}}},\ }\href {\doibase 10.1038/nature08609} {\bibfield
   {journal} {\bibinfo  {journal} {Nature}\ }\textbf {\bibinfo {volume}
  {462}},\ \bibinfo {pages} {628} (\bibinfo {year}
  {2009}{\natexlab{b}})}\BibitemShut {NoStop}%
\bibitem [{\citenamefont {{Simon}}\ \emph {et~al.}(2011)\citenamefont
  {{Simon}}, \citenamefont {{Bakr}}, \citenamefont {{Ma}}, \citenamefont
  {{Tai}}, \citenamefont {{Preiss}},\ and\ \citenamefont
  {{Greiner}}}]{simon11}%
  \BibitemOpen
  \bibfield  {author} {\bibinfo {author} {\bibfnamefont {J.}~\bibnamefont
  {{Simon}}}, \bibinfo {author} {\bibfnamefont {W.~S.}\ \bibnamefont {{Bakr}}},
  \bibinfo {author} {\bibfnamefont {R.}~\bibnamefont {{Ma}}}, \bibinfo {author}
  {\bibfnamefont {M.~E.}\ \bibnamefont {{Tai}}}, \bibinfo {author}
  {\bibfnamefont {P.~M.}\ \bibnamefont {{Preiss}}}, \ and\ \bibinfo {author}
  {\bibfnamefont {M.}~\bibnamefont {{Greiner}}},\ }\href@noop {} {\bibfield
  {journal} {\bibinfo  {journal} {Nature}\ }\textbf {\bibinfo {volume} {472}},\
  \bibinfo {pages} {307} (\bibinfo {year} {2011})}\BibitemShut {NoStop}%
\bibitem [{\citenamefont {{Spielman}}(2011)}]{spiel11}%
  \BibitemOpen
  \bibfield  {author} {\bibinfo {author} {\bibfnamefont {I.~B.}\ \bibnamefont
  {{Spielman}}},\ }\href@noop {} {\bibfield  {journal} {\bibinfo  {journal}
  {Nature}\ }\textbf {\bibinfo {volume} {472}},\ \bibinfo {pages} {301}
  (\bibinfo {year} {2011})}\BibitemShut {NoStop}%
\bibitem [{\citenamefont {{Dalibard}}\ \emph {et~al.}(2011)\citenamefont
  {{Dalibard}}, \citenamefont {{Gerbier}}, \citenamefont {{Juzeli{\= u}nas}},\
  and\ \citenamefont {{{\"O}hberg}}}]{dal10b}%
  \BibitemOpen
  \bibfield  {author} {\bibinfo {author} {\bibfnamefont {J.}~\bibnamefont
  {{Dalibard}}}, \bibinfo {author} {\bibfnamefont {F.}~\bibnamefont
  {{Gerbier}}}, \bibinfo {author} {\bibfnamefont {G.}~\bibnamefont {{Juzeli{\=
  u}nas}}}, \ and\ \bibinfo {author} {\bibfnamefont {P.}~\bibnamefont
  {{{\"O}hberg}}},\ }\href@noop {} {\bibfield  {journal} {\bibinfo  {journal}
  {Reviews of Modern Physics}\ }\textbf {\bibinfo {volume} {83}},\ \bibinfo
  {pages} {1523} (\bibinfo {year} {2011})}\BibitemShut {NoStop}%
\bibitem [{\citenamefont {{Juzeli{\= u}nas}}\ \emph {et~al.}(2010)\citenamefont
  {{Juzeli{\= u}nas}}, \citenamefont {{Ruseckas}},\ and\ \citenamefont
  {{Dalibard}}}]{dal10a}%
  \BibitemOpen
  \bibfield  {author} {\bibinfo {author} {\bibfnamefont {G.}~\bibnamefont
  {{Juzeli{\= u}nas}}}, \bibinfo {author} {\bibfnamefont {J.}~\bibnamefont
  {{Ruseckas}}}, \ and\ \bibinfo {author} {\bibfnamefont {J.}~\bibnamefont
  {{Dalibard}}},\ }\href {\doibase 10.1103/PhysRevA.81.053403} {\bibfield
  {journal} {\bibinfo  {journal} {Phys. Rev. A.}\ }\textbf {\bibinfo {volume}
  {81}},\ \bibinfo {pages} {053403} (\bibinfo {year} {2010})}\BibitemShut
  {NoStop}%
\bibitem [{\citenamefont {Juli\'a-D\'iaz}\ \emph {et~al.}(2011)\citenamefont
  {Juli\'a-D\'iaz}, \citenamefont {Dagnino}, \citenamefont {G\"unter},
  \citenamefont {Gra\ss{}}, \citenamefont {Barber\'an}, \citenamefont
  {Lewenstein},\ and\ \citenamefont {Dalibard}}]{diaz11}%
  \BibitemOpen
  \bibfield  {author} {\bibinfo {author} {\bibfnamefont {B.}~\bibnamefont
  {Juli\'a-D\'iaz}}, \bibinfo {author} {\bibfnamefont {D.}~\bibnamefont
  {Dagnino}}, \bibinfo {author} {\bibfnamefont {K.~J.}\ \bibnamefont
  {G\"unter}}, \bibinfo {author} {\bibfnamefont {T.}~\bibnamefont {Gra\ss{}}},
  \bibinfo {author} {\bibfnamefont {N.}~\bibnamefont {Barber\'an}}, \bibinfo
  {author} {\bibfnamefont {M.}~\bibnamefont {Lewenstein}}, \ and\ \bibinfo
  {author} {\bibfnamefont {J.}~\bibnamefont {Dalibard}},\ }\href {\doibase
  10.1103/PhysRevA.84.053605} {\bibfield  {journal} {\bibinfo  {journal} {Phys.
  Rev. A}\ }\textbf {\bibinfo {volume} {84}},\ \bibinfo {pages} {053605}
  (\bibinfo {year} {2011})}\BibitemShut {NoStop}%
\bibitem [{\citenamefont {Mead}\ and\ \citenamefont {Truhlar}(1979)}]{mead76}%
  \BibitemOpen
  \bibfield  {author} {\bibinfo {author} {\bibfnamefont {C.~A.}\ \bibnamefont
  {Mead}}\ and\ \bibinfo {author} {\bibfnamefont {G.~D.}\ \bibnamefont
  {Truhlar}},\ }\href {\doibase 10.1063/1.437734} {\bibfield  {journal}
  {\bibinfo  {journal} {The Journal of Chemical Physics}\ }\textbf {\bibinfo
  {volume} {70}},\ \bibinfo {pages} {2284} (\bibinfo {year}
  {1979})}\BibitemShut {NoStop}%
\bibitem [{\citenamefont {Berry}(1984)}]{ber84}%
  \BibitemOpen
  \bibfield  {author} {\bibinfo {author} {\bibfnamefont {M.~V.}\ \bibnamefont
  {Berry}},\ }\href@noop {} {\bibfield  {journal} {\bibinfo  {journal} {Proc.
  R. Soc. Lond. A}\ }\textbf {\bibinfo {volume} {392}},\ \bibinfo {pages} {45}
  (\bibinfo {year} {1984})}\BibitemShut {NoStop}%
\bibitem [{\citenamefont {Wilczek}\ and\ \citenamefont {Zee}(1984)}]{zee}%
  \BibitemOpen
  \bibfield  {author} {\bibinfo {author} {\bibfnamefont {F.}~\bibnamefont
  {Wilczek}}\ and\ \bibinfo {author} {\bibfnamefont {A.}~\bibnamefont {Zee}},\
  }\href {\doibase 10.1103/PhysRevLett.52.2111} {\bibfield  {journal} {\bibinfo
   {journal} {Phys. Rev. Lett.}\ }\textbf {\bibinfo {volume} {52}},\ \bibinfo
  {pages} {2111} (\bibinfo {year} {1984})}\BibitemShut {NoStop}%
\bibitem [{\citenamefont {Zygelman}\ \emph {et~al.}(1992)\citenamefont
  {Zygelman}, \citenamefont {Cooper}, \citenamefont {Ford}, \citenamefont
  {Dalgarno}, \citenamefont {Gerratt},\ and\ \citenamefont {Raimondi}}]{zyg92}%
  \BibitemOpen
  \bibfield  {author} {\bibinfo {author} {\bibfnamefont {B.}~\bibnamefont
  {Zygelman}}, \bibinfo {author} {\bibfnamefont {D.~L.}\ \bibnamefont
  {Cooper}}, \bibinfo {author} {\bibfnamefont {M.~J.}\ \bibnamefont {Ford}},
  \bibinfo {author} {\bibfnamefont {A.}~\bibnamefont {Dalgarno}}, \bibinfo
  {author} {\bibfnamefont {J.}~\bibnamefont {Gerratt}}, \ and\ \bibinfo
  {author} {\bibfnamefont {M.}~\bibnamefont {Raimondi}},\ }\href {\doibase
  10.1103/PhysRevA.46.3846} {\bibfield  {journal} {\bibinfo  {journal} {Phys.
  Rev. A}\ }\textbf {\bibinfo {volume} {46}},\ \bibinfo {pages} {3846}
  (\bibinfo {year} {1992})}\BibitemShut {NoStop}%
\bibitem [{\citenamefont {Zygelman}\ \emph {et~al.}(1994)\citenamefont
  {Zygelman}, \citenamefont {Dalgarno},\ and\ \citenamefont {Sharma}}]{zyg94a}%
  \BibitemOpen
  \bibfield  {author} {\bibinfo {author} {\bibfnamefont {B.}~\bibnamefont
  {Zygelman}}, \bibinfo {author} {\bibfnamefont {A.}~\bibnamefont {Dalgarno}},
  \ and\ \bibinfo {author} {\bibfnamefont {R.}~\bibnamefont {Sharma}},\
  }\href@noop {} {\bibfield  {journal} {\bibinfo  {journal} {Phys. Rev. A}\
  }\textbf {\bibinfo {volume} {49}},\ \bibinfo {pages} {2587} (\bibinfo {year}
  {1994})}\BibitemShut {NoStop}%
\bibitem [{\citenamefont {Zygelman}(2009)}]{zyg09}%
  \BibitemOpen
  \bibfield  {author} {\bibinfo {author} {\bibfnamefont {B.}~\bibnamefont
  {Zygelman}},\ }in\ \href@noop {} {\emph {\bibinfo {booktitle} {Proceedings of
  the Dalgarno Celebratory Symposium: Contributions to Atomic, Molecular, and
  Optical Physics, Astrophysics, and Atmospheric Physics}}},\ \bibinfo {editor}
  {edited by\ \bibinfo {editor} {\bibfnamefont {J.~F.}\ \bibnamefont {Babb}},
  \bibinfo {editor} {\bibfnamefont {K.}~\bibnamefont {Kirby}}, and \bibinfo
  {editor} {\bibfnamefont {H.}~\bibnamefont {Sadeghpour}}}\ (\bibinfo  {publisher}
  {World Scientific Publishing Company},\ \bibinfo {year} {2009}) \BibitemShut {NoStop}%
\bibitem [{\citenamefont {Mott}\ and\ \citenamefont {Massey}(1949)}]{mott49}%
  \BibitemOpen
  \bibfield  {author} {\bibinfo {author} {\bibfnamefont {N.~F.}\ \bibnamefont
  {Mott}}\ and\ \bibinfo {author} {\bibfnamefont {H.~S.~W.}\ \bibnamefont
  {Massey}},\ }\href@noop {} {\emph {\bibinfo {title} {The Theory of Atomic
  Collisions}}},\ \bibinfo {edition} {3rd}\ ed.\ (\bibinfo  {publisher}
  {Oxford},\ \bibinfo {year} {1965})\ p.\ \bibinfo {pages} {428}\BibitemShut
  {NoStop}%
\bibitem [{\citenamefont {Smith}(1969)}]{Smith69}%
  \BibitemOpen
  \bibfield  {author} {\bibinfo {author} {\bibfnamefont {F.~T.}\ \bibnamefont
  {Smith}},\ }\href {\doibase 10.1103/PhysRev.179.111} {\bibfield  {journal}
  {\bibinfo  {journal} {Phys. Rev.}\ }\textbf {\bibinfo {volume} {179}},\
  \bibinfo {pages} {111} (\bibinfo {year} {1969})}\BibitemShut {NoStop}%
\bibitem [{\citenamefont {Berry}\ and\ \citenamefont {Robbins}(1992)}]{ber93}%
  \BibitemOpen
  \bibfield  {author} {\bibinfo {author} {\bibfnamefont {M.~V.}\ \bibnamefont
  {Berry}}\ and\ \bibinfo {author} {\bibfnamefont {J.~M.}\ \bibnamefont
  {Robbins}},\ }\href@noop {} {\bibfield  {journal} {\bibinfo  {journal} {Proc.
  R. Soc. Lond.}\ }\textbf {\bibinfo {volume} {A442}},\ \bibinfo {pages} {641}
  (\bibinfo {year} {1992})}\BibitemShut {NoStop}%
\bibitem [{\citenamefont {Berry}(1996)}]{ber96}%
  \BibitemOpen
  \bibfield  {author} {\bibinfo {author} {\bibfnamefont {M.~V.}\ \bibnamefont
  {Berry}},\ }\href@noop {} {\bibfield  {journal} {\bibinfo  {journal} {Proc.
  R. Soc. Lond. A}\ }\textbf {\bibinfo {volume} {452}},\ \bibinfo {pages}
  {1207} (\bibinfo {year} {1996})}\BibitemShut {NoStop}%
\bibitem [{\citenamefont {Berry}\ and\ \citenamefont {Shukla}(2010)}]{berry10}%
  \BibitemOpen
  \bibfield  {author} {\bibinfo {author} {\bibfnamefont {M.~V.}\ \bibnamefont
  {Berry}}\ and\ \bibinfo {author} {\bibfnamefont {P.}~\bibnamefont {Shukla}},\
  }\href@noop {} {\bibfield  {journal} {\bibinfo  {journal} {J. Phys. A: Math.
  Theor.}\ }\textbf {\bibinfo {volume} {43}},\ \bibinfo {pages} {045102}
  (\bibinfo {year} {2010})}\BibitemShut {NoStop}%
\bibitem{mansour2000}
Robert L. Karp, Freydoon Mansouri, Jung S. Rno, Turk. J. Phy. {\bf 24}, 365, (2000).
\bibitem [{\citenamefont {{Wu}}\ and\ \citenamefont {{Yang}}(1976)}]{wuyang76}%
  \BibitemOpen
  \bibfield  {author} {\bibinfo {author} {\bibfnamefont {T.~T.}\ \bibnamefont
  {{Wu}}}\ and\ \bibinfo {author} {\bibfnamefont {C.~N.}\ \bibnamefont
  {{Yang}}},\ }\href {\doibase 10.1016/0550-3213(76)90143-7} {\bibfield
  {journal} {\bibinfo  {journal} {Nuclear Physics B}\ }\textbf {\bibinfo
  {volume} {107}},\ \bibinfo {pages} {365} (\bibinfo {year}
  {1976})}\BibitemShut {NoStop}%
%\bibitem [{Note2()}]{Note2}%
%  \BibitemOpen
%  \bibinfo {note} {For the sake of simplicity, we adopted this, not the most
%  general choice.}\BibitemShut {Stop}%
\bibitem [{\citenamefont {Ko\l{}os}\ and\ \citenamefont
  {Wolniewicz}(1963)}]{kolos}%
  \BibitemOpen
  \bibfield  {author} {\bibinfo {author} {\bibfnamefont {W.}~\bibnamefont
  {Ko\l{}os}}\ and\ \bibinfo {author} {\bibfnamefont {L.}~\bibnamefont
  {Wolniewicz}},\ }\href {\doibase 10.1103/RevModPhys.35.473} {\bibfield
  {journal} {\bibinfo  {journal} {Rev. Mod. Phys.}\ }\textbf {\bibinfo {volume}
  {35}},\ \bibinfo {pages} {473} (\bibinfo {year} {1963})}\BibitemShut
  {NoStop}%
\bibitem [{\citenamefont {Marinescu}\ and\ \citenamefont
  {Dalgarno}(1998)}]{mari98}%
  \BibitemOpen
  \bibfield  {author} {\bibinfo {author} {\bibfnamefont {M.}~\bibnamefont
  {Marinescu}}\ and\ \bibinfo {author} {\bibfnamefont {A.}~\bibnamefont
  {Dalgarno}},\ }\href@noop {} {\bibfield  {journal} {\bibinfo  {journal}
  {Phys. Rev. A}\ }\textbf {\bibinfo {volume} {57}},\ \bibinfo {pages} {1821}
  (\bibinfo {year} {1998})}\BibitemShut {NoStop}%
\bibitem [{\citenamefont {Gosselin}\ and\ \citenamefont
  {Mohrbach}(2010)}]{goss2010}%
  \BibitemOpen
  \bibfield  {author} {\bibinfo {author} {\bibfnamefont {P.}~\bibnamefont
  {Gosselin}}\ and\ \bibinfo {author} {\bibfnamefont {H.}~\bibnamefont
  {Mohrbach}},\ }\href@noop {} {\bibfield  {journal} {\bibinfo  {journal} {J.
  Phys. A: Math. Theor.}\ }\textbf {\bibinfo {volume} {43}},\ \bibinfo {pages}
  {354025} (\bibinfo {year} {2010})}\BibitemShut {NoStop}%
\bibitem [{\citenamefont {{Feshbach}}(1958)}]{fesh58}%
  \BibitemOpen
  \bibfield  {author} {\bibinfo {author} {\bibfnamefont {H.}~\bibnamefont
  {{Feshbach}}},\ }\href {\doibase 10.1016/0003-4916(58)90007-1} {\bibfield
  {journal} {\bibinfo  {journal} {Annals of Physics}\ }\textbf {\bibinfo
  {volume} {5}},\ \bibinfo {pages} {357} (\bibinfo {year} {1958})}\BibitemShut
  {NoStop}%
\bibitem [{\citenamefont {Zygelman}(2011)}]{zyg11a}%
  \BibitemOpen
  \bibfield  {author} {\bibinfo {author} {\bibfnamefont {B.}~\bibnamefont
  {Zygelman}},\ }\href@noop {} {\bibfield  {journal} {\bibinfo  {journal}
  {Unpublished}\ } (\bibinfo {year} {2011})}\BibitemShut {NoStop}%
\bibitem [{\citenamefont {{Juzeli{\= u}nas}}\ \emph {et~al.}(2006)\citenamefont
  {{Juzeli{\= u}nas}}, \citenamefont {{Ruseckas}}, \citenamefont
  {{{\"O}hberg}},\ and\ \citenamefont {{Fleischhauer}}}]{juz06}%
  \BibitemOpen
  \bibfield  {author} {\bibinfo {author} {\bibfnamefont {G.}~\bibnamefont
  {{Juzeli{\= u}nas}}}, \bibinfo {author} {\bibfnamefont {J.}~\bibnamefont
  {{Ruseckas}}}, \bibinfo {author} {\bibfnamefont {P.}~\bibnamefont
  {{{\"O}hberg}}}, \ and\ \bibinfo {author} {\bibfnamefont {M.}~\bibnamefont
  {{Fleischhauer}}},\ }\href {\doibase 10.1103/PhysRevA.73.025602} {\bibfield
  {journal} {\bibinfo  {journal} {Phys. Rev. A.}\ }\textbf {\bibinfo {volume}
  {73}},\ \bibinfo {pages} {025602} (\bibinfo {year} {2006})}\BibitemShut
  {NoStop}%
%\bibitem [{Note3()}]{Note3}%
%  \BibitemOpen
%  \bibinfo {note} {When we use the term exact, we imply that a sufficiently
%  accurate algorithm exists in which numerical solution of the coupled
%  equations can be obtained to the desired level of accuracy.}\BibitemShut
%  {Stop}%
\bibitem [{\citenamefont {Nambu}\ and\ \citenamefont
  {Jona-Lasino}(1961)}]{nambu61}%
  \BibitemOpen
  \bibfield  {author} {\bibinfo {author} {\bibfnamefont {Y.}~\bibnamefont
  {Nambu}}\ and\ \bibinfo {author} {\bibfnamefont {G.}~\bibnamefont
  {Jona-Lasino}},\ }\href@noop {} {\bibfield  {journal} {\bibinfo  {journal}
  {Physical Review}\ }\textbf {\bibinfo {volume} {122}},\ \bibinfo {pages}
  {345} (\bibinfo {year} {1961})}\BibitemShut {NoStop}%
\bibitem [{\citenamefont {{Hermann}}\ and\ \citenamefont
  {{Fleck}}(1988)}]{fleck88}%
  \BibitemOpen
  \bibfield  {author} {\bibinfo {author} {\bibfnamefont {M.~R.}\ \bibnamefont
  {{Hermann}}}\ and\ \bibinfo {author} {\bibfnamefont {J.~A.}\ \bibnamefont
  {{Fleck}}, \bibfnamefont {Jr.}},\ }\href {\doibase 10.1103/PhysRevA.38.6000}
  {\bibfield  {journal} {\bibinfo  {journal} {\pra}\ }\textbf {\bibinfo
  {volume} {38}},\ \bibinfo {pages} {6000} (\bibinfo {year}
  {1988})}\BibitemShut {NoStop}%
\bibitem [{\citenamefont {{Zygelman}}(2010{\natexlab{a}})}]{zyg10d}%
  \BibitemOpen
  \bibfield  {author} {\bibinfo {author} {\bibfnamefont {B.}~\bibnamefont
  {{Zygelman}}},\ }\href@noop {} {\bibfield  {journal} {\bibinfo  {journal}
  {APS Meeting Abstracts}\ ,\ \bibinfo {pages} {1027}} (\bibinfo {year}
  {2010}{\natexlab{a}})}\BibitemShut {NoStop}%
\bibitem [{\citenamefont {Aharonov}\ and\ \citenamefont {Bohm}(1959)}]{ab59}%
  \BibitemOpen
  \bibfield  {author} {\bibinfo {author} {\bibfnamefont {A.}~\bibnamefont
  {Aharonov}}\ and\ \bibinfo {author} {\bibfnamefont {D.}~\bibnamefont
  {Bohm}},\ }\href@noop {} {\bibfield  {journal} {\bibinfo  {journal} {Phys.
  Rev.}\ }\textbf {\bibinfo {volume} {115}},\ \bibinfo {pages} {485} (\bibinfo
  {year} {1959})}\BibitemShut {NoStop}%
\bibitem [{\citenamefont {Mead}(1992)}]{mead92}%
  \BibitemOpen
  \bibfield  {author} {\bibinfo {author} {\bibfnamefont {C.~A.}\ \bibnamefont
  {Mead}},\ }\href {\doibase 10.1103/RevModPhys.64.51} {\bibfield  {journal}
  {\bibinfo  {journal} {Rev. Mod. Phys.}\ }\textbf {\bibinfo {volume} {64}},\
  \bibinfo {pages} {51} (\bibinfo {year} {1992})}\BibitemShut {NoStop}%
\bibitem [{\citenamefont {{Juanes-Marcos}}\ \emph {et~al.}(2005)\citenamefont
  {{Juanes-Marcos}}, \citenamefont {{Althorpe}},\ and\ \citenamefont
  {{Wrede}}}]{marcos05}%
  \BibitemOpen
  \bibfield  {author} {\bibinfo {author} {\bibfnamefont {J.~C.}\ \bibnamefont
  {{Juanes-Marcos}}}, \bibinfo {author} {\bibfnamefont {S.~C.}\ \bibnamefont
  {{Althorpe}}}, \ and\ \bibinfo {author} {\bibfnamefont {E.}~\bibnamefont
  {{Wrede}}},\ }\href {\doibase 10.1126/science.1114890} {\bibfield  {journal}
  {\bibinfo  {journal} {Science}\ }\textbf {\bibinfo {volume} {309}},\ \bibinfo
  {pages} {1227} (\bibinfo {year} {2005})}\BibitemShut {NoStop}%
\bibitem [{\citenamefont {Stone}(1986)}]{stone86}%
  \BibitemOpen
  \bibfield  {author} {\bibinfo {author} {\bibfnamefont {M.}~\bibnamefont
  {Stone}},\ }\href {\doibase 10.1103/PhysRevD.33.1191} {\bibfield  {journal}
  {\bibinfo  {journal} {Phys. Rev. D}\ }\textbf {\bibinfo {volume} {33}},\
  \bibinfo {pages} {1191} (\bibinfo {year} {1986})}\BibitemShut {NoStop}%
\bibitem [{\citenamefont {{Dirac}}(1931)}]{dirac31}%
  \BibitemOpen
  \bibfield  {author} {\bibinfo {author} {\bibfnamefont {P.~A.~M.}\
  \bibnamefont {{Dirac}}},\ }\href@noop {} {\bibfield  {journal} {\bibinfo
  {journal} {Royal Society of London Proceedings Series A}\ }\textbf {\bibinfo
  {volume} {133}},\ \bibinfo {pages} {60} (\bibinfo {year} {1931})}\BibitemShut
  {NoStop}%
\bibitem{russel}
John March-Russel, John Preskill, Frank Wilczek, Phys. Rev. Lett. {\bf 68}, 2567 (1992).
\bibitem [{\citenamefont {{Zygelman}}(2010{\natexlab{b}})}]{zyg10b}%
  \BibitemOpen
  \bibfield  {author} {\bibinfo {author} {\bibfnamefont {B.}~\bibnamefont
  {{Zygelman}}},\ }\href@noop {} {\bibfield  {journal} {\bibinfo  {journal}
  {APS Meeting Abstracts}\ ,\ \bibinfo {pages} {1173}} (\bibinfo {year}
  {2010}{\natexlab{b}})}\BibitemShut {NoStop}%
\bibitem [{\citenamefont {{Zygelman}}(2010{\natexlab{c}})}]{zyg10}%
  \BibitemOpen
  \bibfield  {author} {\bibinfo {author} {\bibfnamefont {B.}~\bibnamefont
  {{Zygelman}}},\ }\href {\doibase 10.1103/PhysRevA.81.032506} {\bibfield
  {journal} {\bibinfo  {journal} {Phys. Rev. A.}\ }\textbf {\bibinfo {volume}
  {81}},\ \bibinfo {pages} {032506} (\bibinfo {year}
  {2010}{\natexlab{c}})}\BibitemShut {NoStop}%
\bibitem [{\citenamefont {Carpenter}(1966)}]{carpenter}%
  \BibitemOpen
  \bibfield  {author} {\bibinfo {author} {\bibfnamefont {W.}~\bibnamefont
  {Carpenter}},\ }\href@noop {} {\bibfield  {journal} {\bibinfo  {journal}
  {Mathematics Magazine}\ }\textbf {\bibinfo {volume} {39}},\ \bibinfo {pages}
  {28} (\bibinfo {year} {1966})}\BibitemShut {NoStop}%
\bibitem [{\citenamefont {Miller}(1972)}]{abrom}%
  \BibitemOpen
  \bibfield  {author} {\bibinfo {author} {\bibfnamefont {J.~C.~P.}\
  \bibnamefont {Miller}},\ }in\ \href@noop {} {\emph {\bibinfo {booktitle}
  {Handbook of Mathematical Functions}}},\ \bibinfo {editor} {edited by\
  \bibinfo {editor} {\bibfnamefont {M.}~\bibnamefont {Abramowitz}}\ and\
  \bibinfo {editor} {\bibfnamefont {I.~A.}\ \bibnamefont {Stegun}}}\ (\bibinfo
  {publisher} {Dover Publications, Inc. New York},\ \bibinfo {year} {1972})\
  p.\ \bibinfo {pages} {685}\BibitemShut {NoStop}%
\end{thebibliography}


\end{document}